\documentclass[sigconf]{acmart}
\AtBeginDocument{%
  }

\setcopyright{acmlicensed}
\copyrightyear{2026}
\acmYear{2026}
\acmDOI{}
\acmConference[]{UIST 2026}
\acmISBN{}

\usepackage{float}
\usepackage{tikz}
\usepackage{xcolor}
\usepackage{enumitem}
\newcommand{\pill}[1]{%
  \tikz[baseline=(X.base)] 
    \node (X) [
      draw=gray!40,
      fill=gray!10,
      rounded corners=8pt,
      inner xsep=4pt,
      inner ysep=1.5pt,
      font=\scriptsize
    ] {\strut #1};%
}
\begin{document}

\title{Conversational Customization of Productivity Systems: A Design Probe of Malleable AI Interfaces}

%
\author{Karthik Sreedhar}
\email{ks4190@columbia.edu}
\affiliation{%
  \institution{Columbia University}
  \city{New York City}
  \state{New York}
  \country{USA}
}

\author{Aryan Kaul}
\email{ak5638@columbia.edu}
\affiliation{%
  \institution{Columbia University}
  \city{New York City}
  \state{New York}
  \country{USA}
  }

\author{Lydia B. Chilton}
\email{chilton@cs.columbia.edu}
\affiliation{%
  \institution{Columbia University}
  \city{New York City}
  \state{New York}
  \country{USA}
}

\renewcommand{\shortauthors}{}

\begin{abstract}
Customization has long been a central goal in interactive systems, yet prior work shows that end-user tailoring occurs infrequently and is often confined to initial setup or moments of breakdown. Recent advances in generative AI suggest that highly malleable systems—where users can modify system behavior through natural language—are now technically feasible. However, it remains unclear how such malleability is used in practice: What kinds of customizations do users create, when do they choose to customize, and how do these modifications shape their experience of everyday tools? We present a design probe that uses a conversationally customizable email system as an instrument to study how users create and refine functionality within everyday tools. The system allows users to iteratively modify their inbox by restructuring categories, introducing interface elements, and authoring new workflow behaviors directly through natural language interaction. We study how participants create, refine, and use these features over several days within their own email workflows. We find that users’ customizations are often grounded in existing patterns, which they adapt and specialize to fit their needs, rather than generating entirely novel functionality. Malleability changes how users engage with their inbox, shifting it from a fixed interface to a flexible data layer shaped through user-authored features. At the same time, customization introduces new forms of risk, including mis-specified behavior, unintended filtering, and uncertainty around outcomes, which users manage through ongoing oversight and refinement. These findings highlight how conversational customization becomes embedded within everyday interaction, and point toward the need for systems that support iterative refinement, visibility into behavior, and safe experimentation as users shape their own tools.
\end{abstract}



\maketitle

\section{Introduction}

Modern productivity tools increasingly embed AI for tasks like prioritization, categorization, recommendation, and content generation.
Foundation models make it possible to interpret and transform messy work artifacts such as emails, documents, and calendar events.
New integrations and agentic workflows allow users to express intent in natural language and have systems carry out multi-step actions across applications.
This shifts expectations from configuring fixed features to shaping tool behavior in an ongoing, conversational way.
Because productivity work is high-frequency and idiosyncratic, even small improvements in workflow fit can compound into meaningful gains in time and attention.
As a result, the limiting factor is often not model capability, but whether systems can adapt to a user’s evolving way of working.
This creates an opportunity to treat software not as a fixed product, but as a malleable system that users can iteratively make their own.

In practice, productivity work is open-ended, evolves over time, and varies substantially across users and contexts.
Users frequently encounter “almost-right” system behavior, where a small change or missing capability would make the system significantly more useful.
However, existing tools provide little support for lightweight, in-the-moment modification of system behavior.
Most systems expose intelligence as fixed features or pre-defined settings that are configured once and rarely revisited.
Generative AI introduces a different possibility: systems that are continuously modifiable through natural language.
This raises a central question:
What happens when users can directly modify the behavior of the systems they use every day?

Customization has long been a central goal in HCI, with prior work arguing that systems should adapt to users’ evolving practices rather than require users to conform to fixed assumptions.
Research on end-user development, programming-by-demonstration, and tailorable systems emphasizes that meaningful customization must occur during use, not only at design time.
Despite this, widespread end-user tailoring remains rare in everyday productivity tools due to barriers such as scripting requirements, abstract configuration interfaces, and detachment from real workflows.
Recent generative AI systems lower the technical barrier to customization by enabling users to express changes through natural language.
However, most existing systems focus on modifying outputs or presentation, or operate in exploratory environments rather than embedding customization into tools users rely on daily.
As a result, while highly malleable systems are now technically feasible, it remains unclear how users engage with them in practice when customization is embedded directly into use.

To investigate this design space, we conducted a design probe using a conversationally malleable email system.
We chose email because it is a high-frequency, everyday tool that users already rely on extensively, making it a natural setting for observing real customization behavior.
The system presents a representation of a user’s inbox augmented with a conversational agent that allows users to modify system behavior itself, not just its outputs.
Users can create and refine features such as new categories, interface elements, and workflow behaviors through natural language interaction, with changes occurring incrementally and in context.

We worked with eight users over a three-day period, combining guided onboarding with continued use and iterative refinement of user-authored features.
On the first day, users proposed and implemented a feature based on their own workflow needs.
In subsequent sessions, they reflected on use, refined features, explored others’ features, or proposed new ones.
Rather than evaluating a fixed system, we use this probe to study how users shape system behavior through feature creation and use.

We find that when customization is directly accessible through conversational interaction, users actively reshape their tools in response to everyday friction.
Rather than creating entirely novel functionality, participants grounded their features in existing patterns and iteratively adapted them to fit their workflows.
These customizations transformed the inbox from a fixed interface into a flexible data layer, with user-authored features shaping how information is accessed and acted upon.

We identify three key patterns.
First, users’ ideas for customization are largely shaped by existing examples, which they adapt and specialize to their own workflows.
Second, malleability changes how users experience their inbox, shifting it from a fixed interface to a personalized information environment.
Third, user-authored functionality introduces new risks, which users manage through oversight, iteration, and selective trust.

This work provides empirical and design insights into how users create, refine, and evaluate custom functionality in malleable AI systems. Rather than evaluating system performance, this work uses a system as a design probe to surface how users engage with malleable interfaces in practice.
System modification should be treated as a first-class interaction, not an advanced or hidden capability.
Conversational interfaces can function as meta-interfaces that allow users to inspect and control system behavior itself.
As intelligent systems become more powerful, usability hinges not only on what systems can do, but on how easily users can shape them to fit their own workflows.
\section{Related Work}

\subsection{Customization as Ongoing Adaptation}

The idea that software should adapt to users has motivated decades of HCI research. Early work on user-tailorable systems argued that systems must remain modifiable during use and developed button-based tailoring mechanisms that allowed users to encode preferred interaction patterns without programming expertise~\cite{maclean1990}. Subsequent research found that tailoring is uncommon in everyday practice, tending to cluster around moments of initial setup, breakdown, or environmental change~\cite{mackay1991triggers}. A follow-up study two decades later observed that the same barriers persisted, suggesting the problem is not simply technical; interface design itself can either trigger or suppress customization behavior~\cite{banovic2012triggering}.

End-user programming research has explored how to give non-developers the ability to create and modify software artifacts. A comprehensive survey of end-user software engineering documents how people ranging from accountants to scientists write programs in spreadsheets, scripts, and domain-specific tools, often without formal training and with limited support for testing or debugging~\cite{ko2011euse}. Related work argues that end-user programming is pervasive but poorly supported, with most tools requiring users to operate in environments detached from their primary tasks~\cite{myers2006eup}. End users face both technical and cognitive barriers when attempting to customize systems~\cite{ko2011euse}.

The concept of appropriation broadens this picture, framing adaptation as the process by which people bend technologies to fit practices that designers never anticipated~\cite{dourish2003}. Technologies become meaningful through ongoing use, as users discover ways to repurpose tools for their own needs. 

Beyond traditional end-user programming, workflow automation tools such as trigger-action systems (e.g., IFTTT, Zapier) have attempted to lower the barrier to customization by allowing users to define lightweight rules over their data and applications~\cite{ur2016ifttt, zapier2020}. However, these systems require users to anticipate needs in advance and operate outside the flow of everyday use.

Taken together, this body of work suggests that the rarity of end-user tailoring reflects a mismatch between where customization needs arise (in the flow of real work, driven by concrete friction) and where customization mechanisms live (in separate configuration environments, requiring planning and effort that most users never invest)~\cite{mackay1991triggers, banovic2012triggering, dourish2003, ko2011euse}. The distance between feeling a frustration and being able to act on it remains too large.

\subsection{Generative AI and Malleable Interfaces}

Recent advances in generative AI have substantially lowered the technical barrier to customization by enabling users to specify and revise interfaces through natural language. This body of work spans several distinct contributions worth distinguishing.

At the \emph{substrate} level, recent work proposes task-driven data models as a foundation for generative user interfaces, showing that LLMs can produce and maintain interface specifications from high-level user goals~\cite{cao2025}. Generative AI can therefore serve as an ongoing substrate that adapts to changing task requirements, rather than producing a single fixed interface.

Along the dimensions of \emph{content and composition}, malleable overview-detail interfaces allow users to reshape what information is displayed and how it is laid out, for example requesting that a product list appear on a map rather than in a table~\cite{min2025overview}. Meridian formalizes design principles for this class of interface~\cite{min2025meridian}.

At the level of \emph{runtime modification}, MorphGUI demonstrates that real-time LLM-driven customization of GUI properties is feasible through conversational interaction~\cite{calo2025morphgui}. Work on gradual generation takes a complementary approach, arguing that malleable systems need interaction structures that make customization options discoverable over time~\cite{min2025gradual}. This echoes long-standing challenges in end-user programming, where users often do not know what is possible to customize until they encounter a specific need~\cite{ko2011euse, myers2006eup}.

BlendScape highlights a tension central to our own findings: generative customization increases expressive power, but users also need controls to manage unpredictable outputs~\cite{rajaram2024blendscape}. In a study of generative video-conferencing environments, users required mechanisms to constrain and correct system-generated modifications, a challenge that resurfaces in our findings around mis-specified features and iterative refinement.

Together, these systems show that malleable interfaces are technically feasible and increasingly usable. However, most operate within exploratory or prototype environments~\cite{cao2025, min2025overview, min2025gradual} and primarily affect content presentation or layout~\cite{calo2025morphgui, rajaram2024blendscape}. Changing how information looks is not the same as changing how a user works. 

In parallel, a growing class of AI-powered productivity systems and agentic tools enables users to generate content, execute multi-step actions, and interact with software through natural language (e.g., email copilots, task assistants, and LLM-based agents)~\cite{yao2023react, openai2024assistants}. These systems expand what users can do but treat functionality as fixed, limiting user control over system behavior.

Less is known about what happens when users can author new functionality within tools they already depend on, reshaping the workflows and information structures underneath the interface.

\subsection{Persistent Customization in Everyday Tools}

AI is already deeply embedded in email. Smart Reply generates short response suggestions from incoming messages~\cite{kannan2016smartreply}, and Smart Compose offers real-time sentence completions as users type~\cite{chen2019smartcompose}. Beyond AI-assisted composition, email clients also offer rule-based customization mechanisms such as filters, labels, and auto-sorting rules. However, these operate on surface-level message properties (sender, subject, keywords) and must be configured in advance through dedicated user settings interfaces, separate from the user's primary workflow. They do not support the kind of semantic, in-context modification that our system enables, such as creating a new dashboard derived from email content or defining behavior based on the semantic content of messages.

More recent AI integrations in productivity tools extend this paradigm by enabling drafting, summarization, and task execution directly within applications such as email clients and document editors\cite{duetai2023, microsoft365copilot2023}. However, these capabilities remain largely bounded to predefined interactions, where users can invoke AI assistance but cannot modify how that assistance operates over time. These systems show that language models can meaningfully reduce effort in high-frequency communication. But users cannot modify what gets suggested, change how completions behave, or extend the system with new capabilities tailored to their own workflows.

A smaller body of work has begun to address how user modifications can persist and circulate beyond individual sessions. MyWebstrates shows that changes to software artifacts can remain shareable, inspectable, and persistent across users and devices~\cite{klokmose2024}. A broader vision of malleable software argues that modification should become routine, and that current software ecosystems lock users out of meaningful adaptation even as AI-assisted development becomes more capable~\cite{litt2025}.

These contributions point toward systems in which customization is an ongoing, cumulative process rather than a one-time configuration event. But less is known about how conversational customization plays out inside high-frequency productivity tools where users already have established workflows. Productivity tools like email clients are, at their core, databases with fixed interfaces. Users interact with the same underlying information (messages, threads, attachments) but have little ability to reshape how that information is organized, surfaced, or acted upon. Feature requests in these settings emerge from recurring friction, and mis-specified behavior carries real consequences.

We address this gap through a design probe~\cite{hutchinson2003probes}, deploying a conversationally customizable email assistant as an instrument to study how users create and refine functionality within an everyday tool.
\section{System}
Our system is an email assistant that allows users to not only manage their inbox, but also modify and extend the system’s behavior through conversational interaction. [code will be made available, redacted for anonymization]. The system consists of two primary components: a Gmail assistant that supports inbox organization and interaction, and a feature-generation agent that enables users to create new functionality. Our goal is not to present a finalized system, but to study how users engage with and shape malleable AI systems in practice.

The system is implemented as a web application with server-side APIs and per-user persistent storage backed by MongoDB. Both subsystems interoperate through shared internal APIs and a common feature lifecycle model. A central feature registry serves as the control plane for extensibility, storing feature metadata, lifecycle state, deployment references, and user-level enablement.

\subsection{User Interaction Model}
Users interact with the system at two levels: performing standard inbox operations and modifying the system itself (Appendix ~\ref{usermodel}). At the base level, users view and manage emails organized into categories, with the system suggesting categorizations and draft responses. At the meta level, users can interact with a conversational agent to modify system behavior. Users can request structural changes (e.g., new categories), reorganize data, or define new functionality (e.g., interface elements or workflows). These modifications are proposed, reviewed, and then integrated into the system. This establishes a recurring interaction pattern: \pill{user identifies friction} $\rightarrow$ \pill{proposes change} $\rightarrow$ \pill{system generates modification} $\rightarrow$ \pill{user approves} $\rightarrow$ \pill{system updates behavior}.

\subsection{Gmail Assistant}

\begin{figure*}[!t]
\centering
\includegraphics[width=\linewidth]{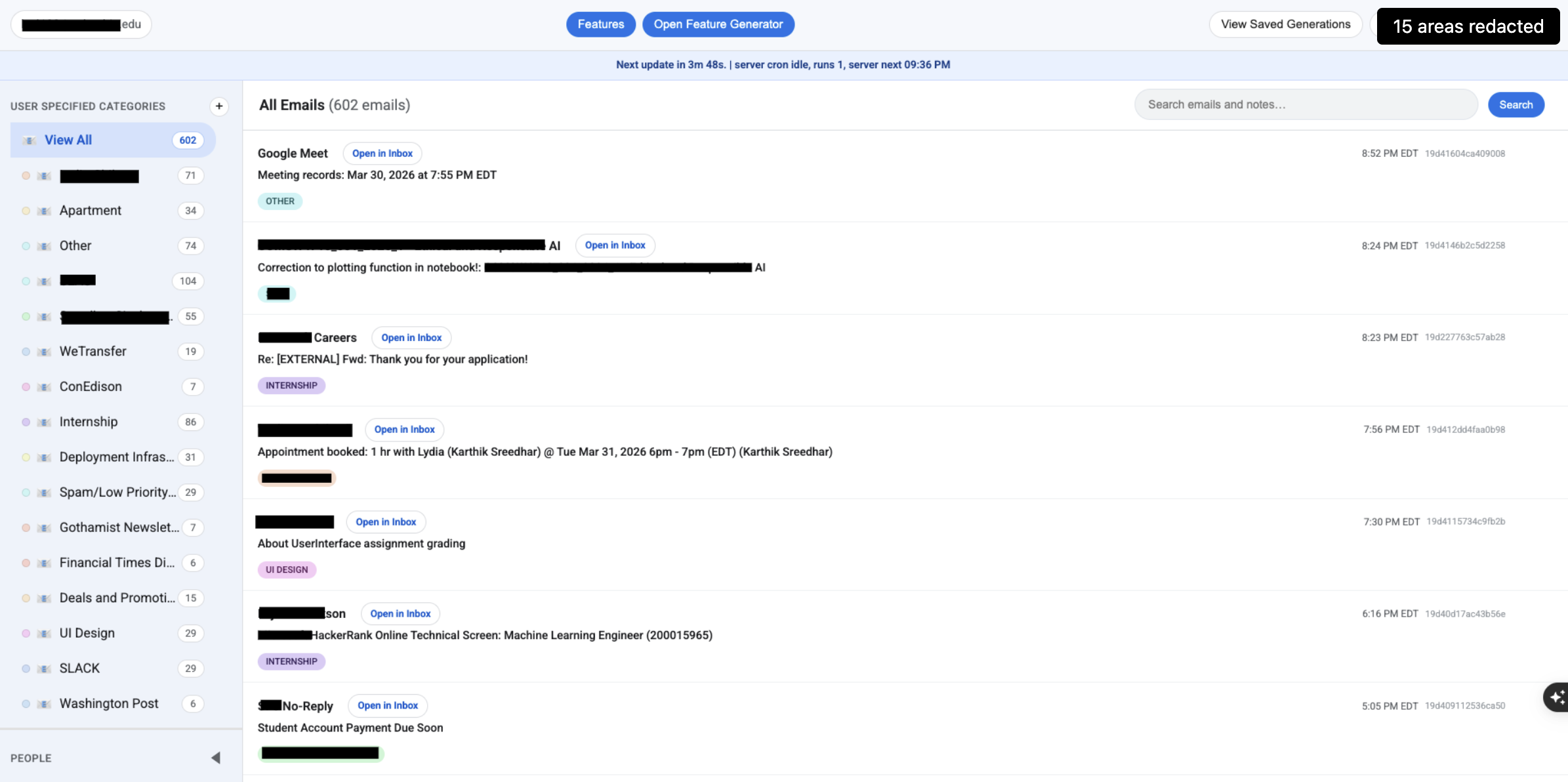}
\caption{
Inbox interface of the email assistant. Emails are organized into user-defined categories in the left sidebar, while the main pane shows categorized email threads, timestamps, and actions such as opening the original message in Gmail. This view serves as the primary workspace for everyday email triage and interaction.
}
\label{fig:gmail-assistant}
\Description{
Screenshot of the main inbox interface of an AI-powered email assistant. A left sidebar lists user-defined categories such as Apartment, Other, Internship, and UI Design, each with a count of emails. The main panel shows a list of email threads with senders, subjects, timestamps, category tags, and buttons to open each message in Gmail. Buttons at the top allow access to features and the feature generator.
}
\end{figure*}

The Gmail assistant manages authentication, inbox synchronization, categorization, and user-facing interaction. Figure~\ref{fig:gmail-assistant} shows the main inbox interface used for day-to-day email interaction.

Upon first login, the system initializes per-user storage and performs a synchronous bootstrap fetch of recent emails (up to 150 messages), ensuring users immediately see populated content. After initialization, new emails are continuously synchronized through a periodic background process (every five minutes). The ingestion pipeline retrieves new messages, de-duplicates them at the thread level, and merges them into normalized per-user stores.

Incoming messages are passed through a categorization pipeline that combines language-model-based classification with user-specific context, including existing categories, summaries, and historical examples. The system produces a suggested category and rationale, which are surfaced to the user and can be left as is or modified. These categorization decisions are stored as editable system state, allowing users to restructure their inbox and enabling downstream customization through the feature-generation agent.

\subsection{Feature Generation Agent}
A conversational agent allows users to interact with the system at a meta-level, enabling them to inspect, modify, and extend its behavior. The agent supports two primary modes: chat and feature generation. This agent is accessible via the "Open Feature Agent" button at the top of the Gmail assistant UI.

\begin{figure*}[!t]
\centering
\includegraphics[width=\linewidth]{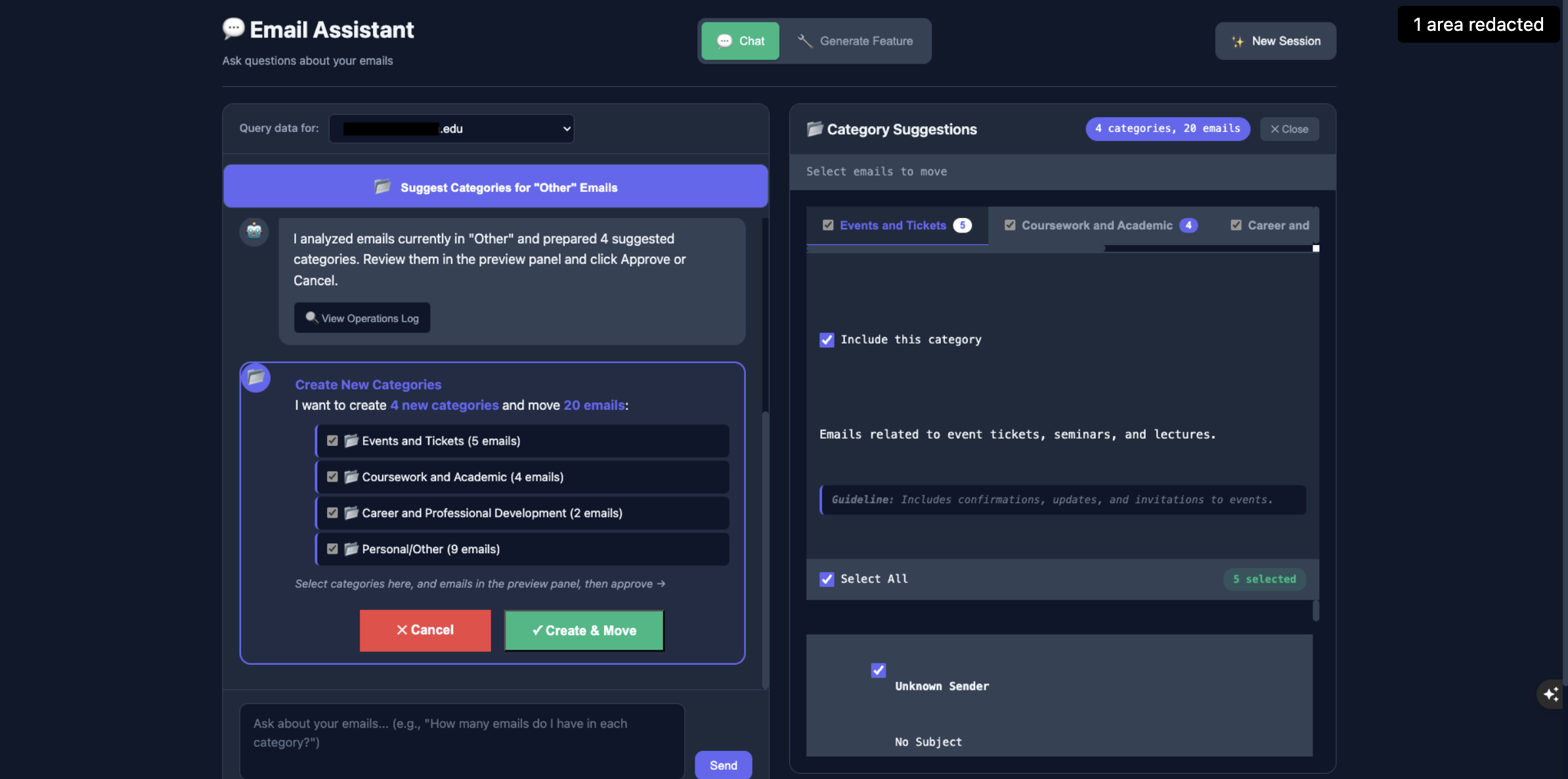}
\caption{
Conversational restructuring interface in chat mode. The agent analyzes emails in ambiguous categories (e.g., ``Other''), proposes new categories, and presents candidate email assignments for user review and approval. This interface supports interactive reorganization of inbox structure through natural language and confirmation-based workflows. The figure demonstrates the outcome of a user asking the system to suggest new categories for their inbox.
}
\label{fig:chat}
\Description{
Screenshot of the chat mode of the email assistant. The left panel shows a conversation where the agent proposes four new categories based on emails currently in Other. The user can select categories to create and move emails into them. The right panel shows detailed category suggestions, selected emails, descriptions, and inclusion controls for reviewing proposed assignments before confirming the changes.
}
\end{figure*}

In chat mode, users can query their inbox, retrieve relevant threads, and request structured modifications to system state, such as adding or removing categories, updating summaries, or reassigning emails (Figure~\ref{fig:chat}). The agent can also analyze emails in ambiguous categories (e.g., “Other”) and propose new category structures with candidate assignments for user approval.

\begin{figure*}[!t]
\centering
\includegraphics[width=\linewidth]{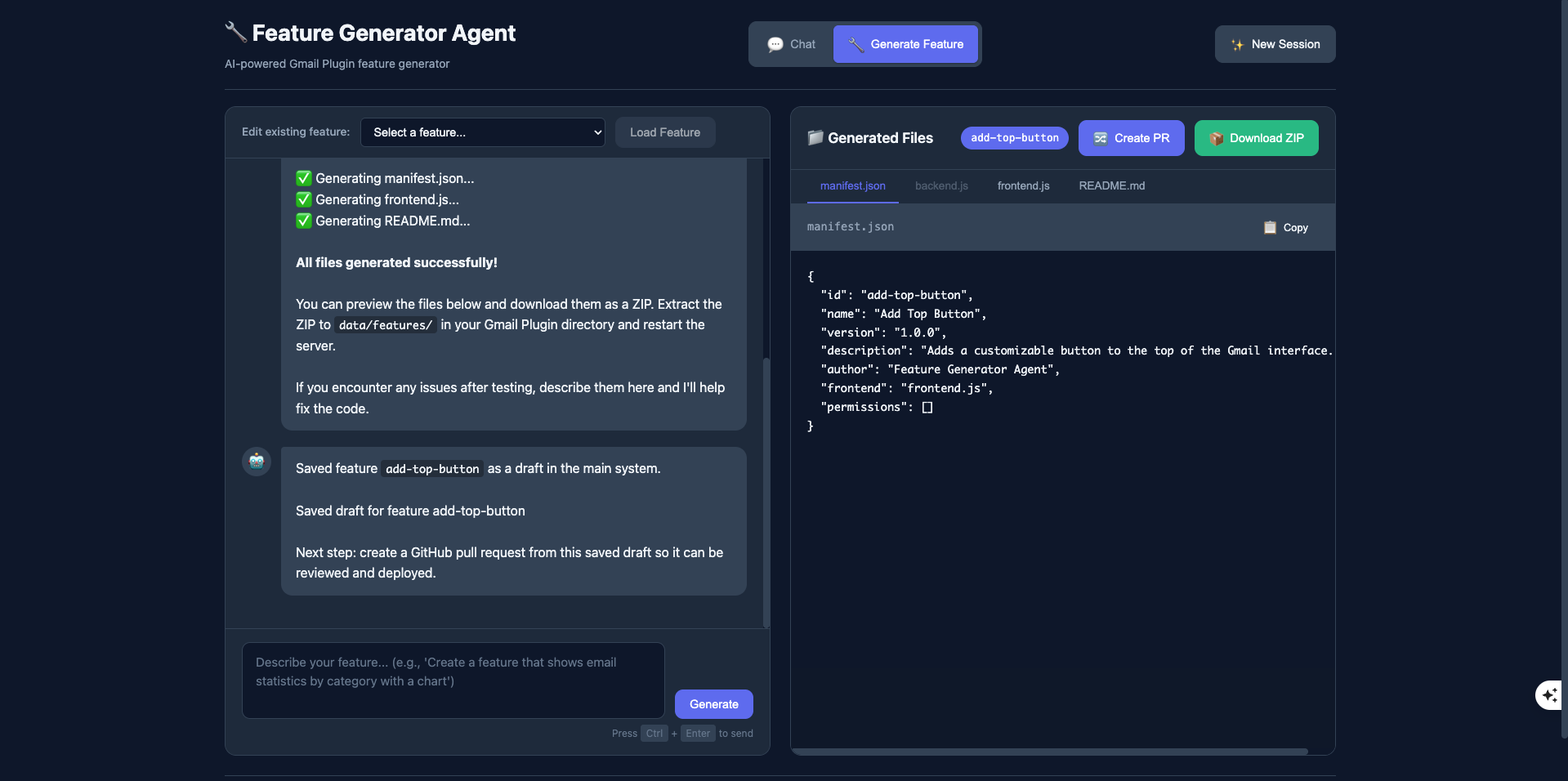}
\caption{
Feature generation interface. Users request new functionality in natural language, inspect generated feature files such as \texttt{manifest.json}, and prepare the result for deployment through a pull-request-based workflow. This interface connects conversational requests to executable, persistent system extensions.
}
\label{fig:generate-feature}
\Description{
Screenshot of the feature generation interface. The left panel shows a conversation confirming that files such as manifest.json, frontend.js, and README.md have been generated successfully and saved as a draft feature. The right panel shows generated files, including the contents of manifest.json, along with buttons for creating a pull request and downloading the generated feature as a ZIP file.
}
\end{figure*}

In feature generation mode, users can request new functionality using natural language (Figure~\ref{fig:generate-feature}). The system translates these requests into executable feature definitions and allows users to iteratively refine them through conversation before installation.

\subsection{Feature Architecture}

To support extensibility, features are implemented as \textbf{standalone modules} rather than modifications to core system logic. Each feature bundle includes metadata (manifest.json), optional backend and frontend components, and documentation.

The system dynamically discovers and loads features at runtime by scanning feature directories and reading their manifests. Features integrate through controlled hooks:
\begin{itemize}
    \item Backend hooks register namespaced API routes (e.g., \newline \texttt{/api/<feature-id>/...})
    \item Frontend hooks attach UI components to the inbox interface
    \item Lifecycle hooks determine visibility and enablement per user
\end{itemize}

This modular structure allows features to be created, modified, or removed without affecting core system functionality. A figure of this architecture is included in Appendix ~\ref{arch}.

\subsection{Feature Lifecycle}

Generated features follow a structured lifecycle from creation to deployment. When a user requests a feature, the system generates a feature bundle containing implementation files and metadata. This bundle is first stored as a draft and can be iteratively refined through conversation.

Once finalized, the system triggers a GitHub-based workflow:
\begin{itemize}
    \item The feature bundle is committed to a new branch
    \item A pull request is created containing the generated files and metadata such as feature ID, user request, and any additional generation context
    \item Upon approval or automated merge, the system is redeployed
\end{itemize}

After deployment, the feature becomes available in the runtime environment and is registered in the feature registry (Appendix ~\ref{app:registry}), which is accessible via the "Features" button at the top of the Gmail assistant UI. Users can enable or disable features, and updates can be made through subsequent iterations that generate new commits. This lifecycle ensures that user-authored features are persistent, auditable (via version history), and automatically deployable (via Git workflows).

\subsection{Design Rationale}

This architecture is designed to treat system modification as a first-class interaction. By enabling users to create features as standalone modules, the system avoids requiring users to understand or modify core logic, lowering the barrier to customization.

The modular feature structure supports safe experimentation. Because features are isolated and explicitly enabled, users can inspect, refine, or disable them without affecting the rest of the system. More broadly, this design allows conversational interaction to function as a meta-interface, enabling users to shape the behavior of the system itself rather than only interacting with its outputs.
\section{Design Probe}
To explore how users engage with conversational customization in a real productivity context, we conducted a design probe in which participants used our system to create and refine features for their own inbox workflows. We designed our probe to investigate how users create, evaluate, and reflect on self-authored features within an AI-mediated system. In particular, we aimed to answer the following research questions:

\begin{itemize}[leftmargin=*]
\item \underline{\textbf{RQ1:}} What gives participants ideas for customizations to create when they have access to a malleable productivity system?
\item \underline{\textbf{RQ2:}} How does malleability change participants’ inbox experiences?
\item \underline{\textbf{RQ3:}} What risks emerge from participant-authored features, and how do participants reason about them?

\color{black}
\end{itemize}

Rather than evaluating a fixed system, we study how participants shape system behavior through feature creation and use. Specifically, we focused on ensuring participants were able to create functional features, stepping in as needed to debug and pivot the feature agent builder. The features created by all participants are included in Appendices ~\ref{designprobe:p0} to ~\ref{designprobe:p7}.

\subsection{Participants and Setup}
We recruited eight participants who regularly use email as part of their daily workflow. The participant group included five men and three women, with backgrounds ranging from computer science PhD students to settlement analysts and law students. This diversity allowed us to observe feature creation across both technical and non-technical domains, as well as across different types of email usage (e.g., academic coordination, professional communication, and administrative tasks). High-level information regarding the participants and the features they created are included in Table ~\ref{tab:participants}.

Each participant interacted with the system over a three-day period, consisting of an initial feature creation session, a follow-up refinement session, and a final survey. Between sessions, participants were instructed to use this system as their default inbox.
Participants used their own email accounts through the system, allowing feature ideas to emerge from real, ongoing tasks rather than hypothetical scenarios.

\begin{table*}[t]
\centering
\small
\begin{tabular}{p{0.10\linewidth} p{0.08\linewidth} p{0.10\linewidth} p{0.26\linewidth} p{0.40\linewidth}}
\toprule
\textbf{Participant} & \textbf{Age} & \textbf{Gender} & \textbf{Occupation} & \textbf{Feature Created} \\
\midrule
P0 & 27 & M & Settlement Analyst & Bloomberg Terminal-Style Newsletter Interface \\
P1 & 26 & M & Data Scientist & Apartment Listings Tracker \\
P2 & 24 & M & Master's Student & Job Application Tracker \\
P3 & 24 & F & PhD Student & Newsletter Interface Highlighting Upcoming Events \\
P4 & 24 & F & PhD Student & Auto-Reply to Students for TA-Related Questions \\
P5 & 25 & M & PhD Student & Filter for Robotics-Related Event Emails \\
P6 & 24 & M & Law Student & Surface Emails with Immediately Upcoming Deadlines \\
P7 & 24 & F & Master's Students & Interface Displaying all JIRA Tasks \\
\bottomrule
\end{tabular}
\caption{
Participant backgrounds and the features they created during the study. Features span organizational tools, automation, filtering, and new interface constructions derived from inbox data.
}
\label{tab:participants}
\end{table*}

\subsection{Study Procedure}
\subsubsection{Day 1: Feature Ideation and Creation}
On the first day, we introduced participants to the system as an augmented inbox interface that supports both standard email interaction and the ability to build new functionality on top of it. We demonstrated a set of example features (images and video links for all example features are included in Appendices ~\ref{ex1} to ~\ref{ex6}) as well as features created by other participants to illustrate the range of possible customizations. Participants then logged into the system and initialized their inbox representation, interacting with the chat to make categories.

The primary goal of this session was for participants to propose and implement a feature grounded in their own workflow needs. To support ideation, we prompted participants to reflect on their current email practices, including:
what types of emails they receive,
what feels repetitive or time-consuming,
and what they wish their inbox could do automatically. Based on this reflection, participants were asked to describe their motivation for their idea. Participants then described their desired feature in natural language and iteratively refined it through interaction with the feature generation agent. Each prompt and modification was recorded as a discrete iteration, capturing how the feature evolved over time.

While the system supported in-situ feature generation, in cases where implementation stalled due to system limitations, we assisted by routing the participant’s prompts through an external coding agent constrained to modifying feature files. This allowed participants to continue exploring their ideas without being blocked by implementation failures, while preserving their original intent.
At the end of the session, participants were asked to use the system as their primary email interface until the next meeting.

\subsubsection{Day 2: Use and Iteration}
On the second day, we focused on how features held up under actual use. We first asked participants whether they had used their feature and, if so, what aspects worked well or did not behave as expected. For participants who used their feature, we probed: whether the feature matched their original motivation, what was missing or incorrect,
and whether they attempted to modify it independently. Participants then refined their feature through additional prompts, often making targeted adjustments to behavior, scope, or interface integration. The refinements participants made are presented in Table ~\ref{tab:iterations}.

\begin{table*}[t]
\centering
\small
\begin{tabular}{p{0.12\linewidth} p{0.38\linewidth} p{0.42\linewidth}}
\toprule
\textbf{Participant} & \textbf{Feature Made} & \textbf{Day 2 Iteration} \\
\midrule
P0 & Bloomberg Terminal-Style Newsletter Interface & Ticker \& Location Keyword Display and FIltering \\
P1 & Apartment Listings Tracker & Extract Number of Savings from Listing Link \\
P2 & Job Application Tracker & Surface Emails Requiring Action/Follow-Up to Top of Inbox \\
P3 & Newsletter Interface Highlighting Upcoming Events & UI Changes to Mirror Email Interface \\
P4 & Auto-Reply to Students for TA-Related Questions & Identify Students Needing Groups \\
P5 & Filter for Robotics-Related Event Emails & No Iteration \\
P6 & Surface Emails with Immediately Upcoming Deadlines & Fix Deadlines \& Add Time Estimates \\
P7 & Interface Displaying all JIRA Tasks & New Feature: Sort Advisor Emails by Project \\
\bottomrule
\end{tabular}
\caption{
Features created by participants on Day 1 and their corresponding refinements or changes on Day 2. The table highlights how initial feature ideas were iteratively adapted based on real usage.
}
\label{tab:iterations}
\end{table*}


There were some exceptions to refinement. In one case, a participant pivoted from their original idea and instead proposed and implemented a new feature better aligned with their actual usage. In another case, a participant was sufficiently satisfied with their feature and did not desire to create anything else.

Participants were also exposed to features created by others and asked whether they would adopt or adapt them for their own inbox.

\subsubsection{Day 3: Reflection and Survey}
On the third day, we led conversations with each participant probing them on how they used their feature. We particularly asked participants about why their feature was helpful and new workflows that their customization enabled that were previously not possible.

Participants also completed a structured survey reflecting on the feature they built and its role in their workflow. The survey captured multiple dimensions of feature experience, including: how well the feature addressed their original motivation, perceived correctness and reliability, effort required to create and refine it, and whether they expect to continue relying on it over time. Additional questions probed long-term adoption, perceived usefulness, and willingness to further extend or modify the feature. We also examined perceptions of risk and trust, including concerns about whether automated behavior could lead to missed or incorrect actions. Finally, participants reviewed features created by others and indicated which they would consider using, providing a broader view of the design space beyond individual needs.

\subsection{Sample User Experience}
We illustrate system use through one participant’s (P0) feature creation and iteration. Remaining participants' information is contained in the Appendix (Appendices ~\ref{designprobe:p0} to ~\ref{designprobe:p7}).

\subsubsection{Day 1: Feature Ideation and Creation}
\begin{figure}[t]
\centering
\includegraphics[width=\linewidth]{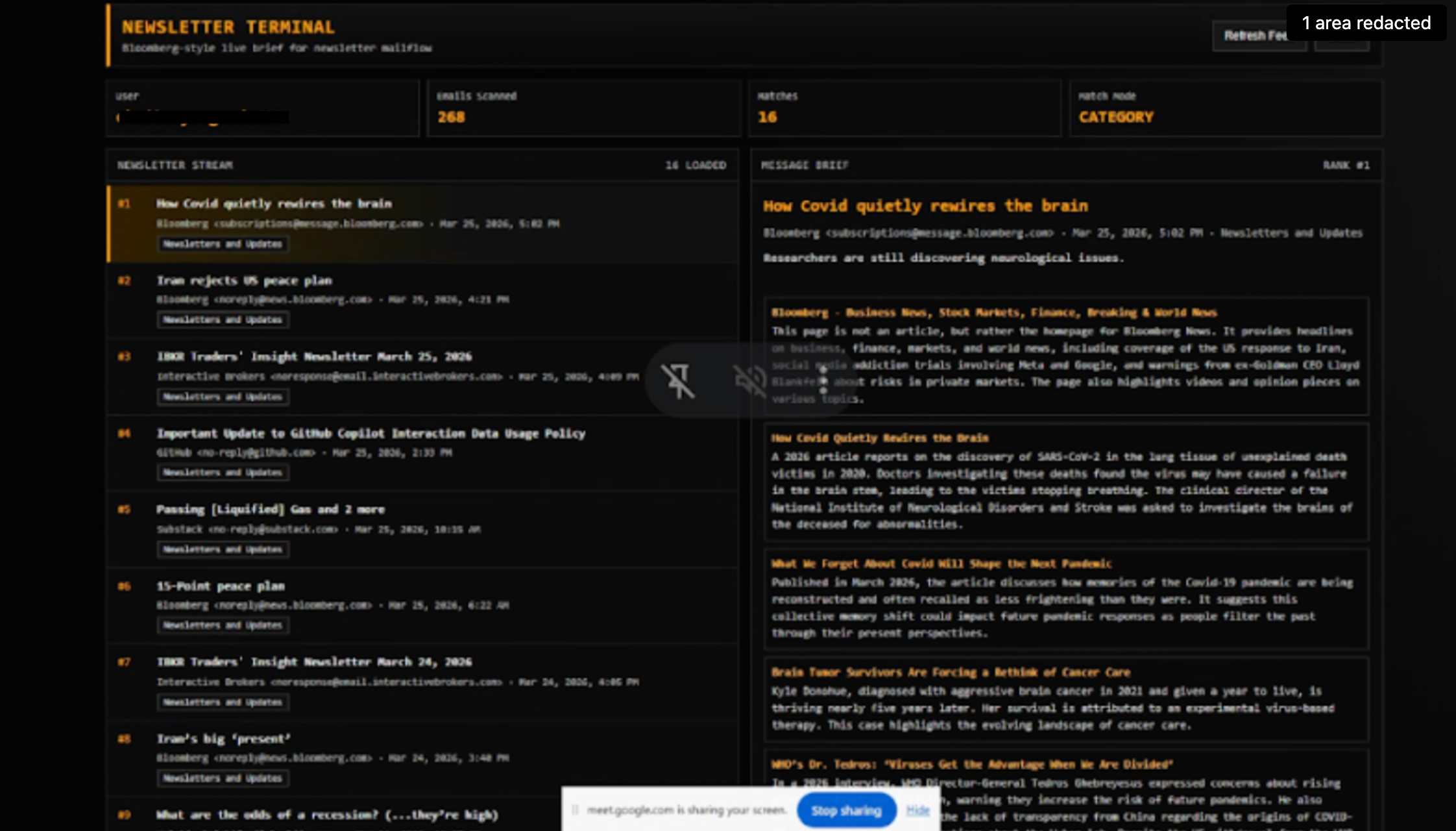}
\caption{
P0’s feature after Day 1: a “Bloomberg terminal-style” newsletter interface that aggregates emails into a standalone view with extracted articles and summaries. A larger view of this figure can be found in  Appendix ~\ref{designprobe:p0}.
}
\label{fig:day1p0}
\end{figure}

P0’s motivation was driven by the many financial newsletters in his inbox. On Day 1, P0 used the feature generation system to request a “Bloomberg terminal-style” interface that would display previews of recent newsletters. His development process then involved multiple refinements. In total, P0 went through four refinements on Day 1, including fixes such as correcting link extraction or UI changes. Through these refinements, the feature evolved from a non-functional UI stub into a working system that extracted, summarized, and displayed newsletter content in a structured interface (Figure ~\ref{fig:day1p0}).


\subsubsection{Day 2: Use and Iteration}
On Day 2, P0 reported actively using the feature and found that it correctly summarized articles, but identified additional needs based on real usage. Specifically, he wanted the system to extract financial tickers and geographic regions mentioned in articles and allow filtering by these attributes.

\begin{figure}[t]
\centering
\includegraphics[width=\linewidth]{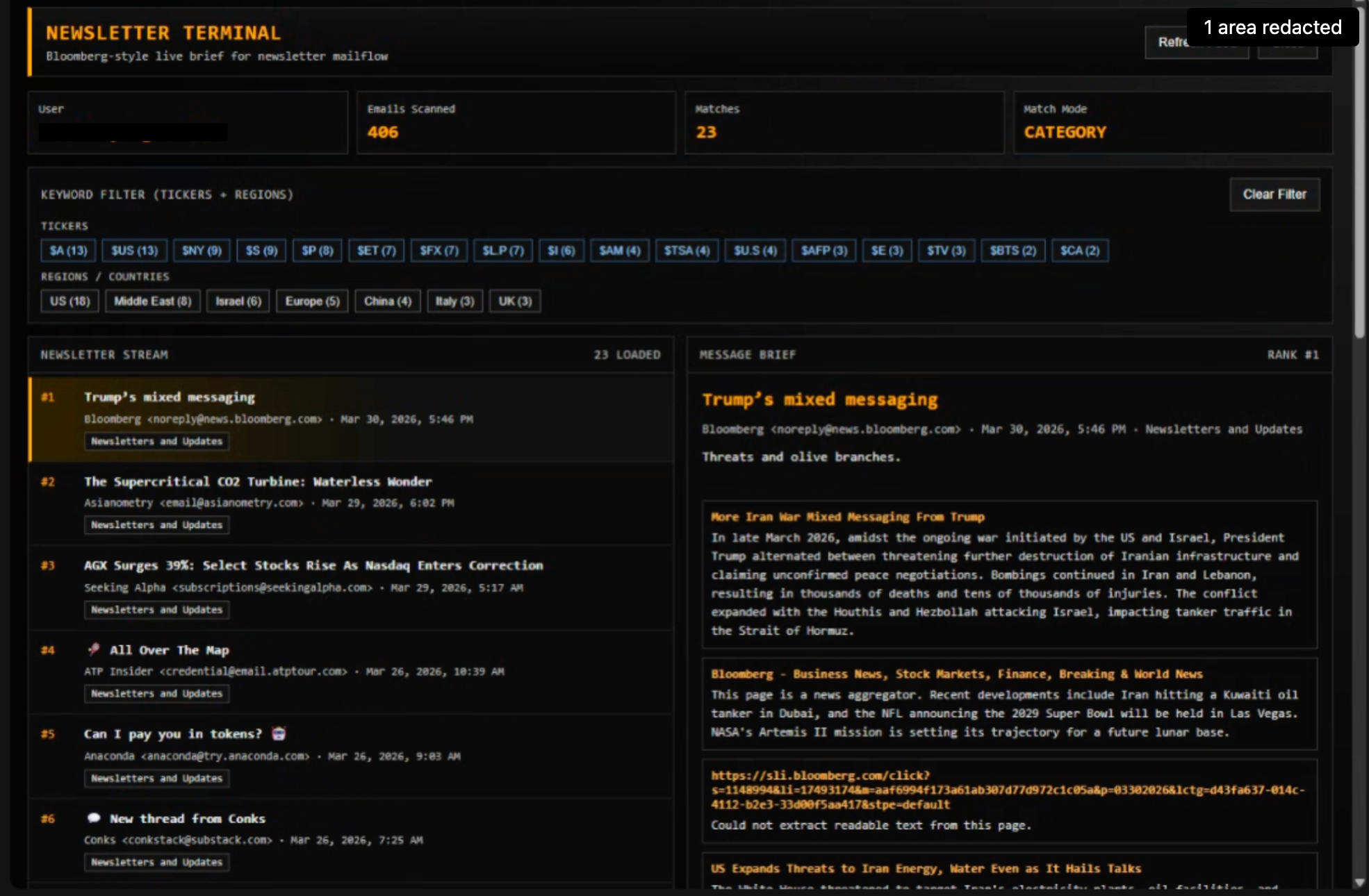}
\caption{
P0’s feature after Day 2: the newsletter interface extended with extracted financial tickers and geographic regions, enabling keyword-based filtering and more targeted exploration of content. A larger view of this figure can be found in Appendix ~\ref{designprobe:p0}.
}
\label{fig:day2p0}
\end{figure}

He proposed and implemented one additional iteration: adding ticker and region metadata with filtering capabilities. The resulting interface (Figure ~\ref{fig:day2p0}) incorporated these changes to the terminal view.

\section{Design Probe Findings}
We present findings on how users engage with and shape a malleable email system through feature creation. All participants were able to create a feature that addressed a meaningful aspect of their email workflow, and reported using it during the study period. Participants’ features aligned closely with their original motivations (mean = 4.38/5), although feature behavior was not always perfectly accurate (mean = 3.75/5). Despite this, participants reported that their features saved time and effort (mean = 3.63/5).

Participants also reported strong understanding of their features (mean = 4.38/5) and a high desire to further extend or refine them (mean = 3.88/5), suggesting that feature creation was not a one-time event but part of an ongoing process. At the same time, users reported moderate comfort in fully trusting these features to run without oversight (mean = 3.88/5), while also expressing relatively low concern about missing important emails (mean = 1.88/5). Together, these results suggest that while users can successfully create and benefit from custom functionality, sustained use depends on how well features align with ongoing needs, integrate into workflows, and maintain reliable behavior over time.

We next present our findings organized by research question.

\subsection{RQ1: What gives participants ideas for customizations to create when they have access to a malleable productivity system?}
Participants’ ideas for customizations were strongly shaped by the demo features presented on the first day (Appendices ~\ref{ex1} to ~\ref{ex6}). While participants found these examples useful and engaging, they often described them as too general for their specific needs. As a result, participants used these features as a \textit{jumping-off point}, adapting and refining them to better fit their own workflows and use cases.

All features created by participants closely resembled one of the demo features. P0, P1, P2, P3, and P7 all created features that closely resembled a FT Digest dashboard shown as a demo feature (Appendix ~\ref{ex6}). P6 created a feature that combined aspects of two features shown in the demo: one which added TODOs onto each email thread (Appendix ~\ref{ex3}), and another which surfaced specific emails to the top of the users inbox (Appendix ~\ref{ex5}). P5 and P7 created filters within an inbox category, closely resembling the filter demo shown (Appendix ~\ref{ex4}). Finally, P4 created an auto-reply feature nearly identical to the quick-reply feature demo shown (Appendix ~\ref{ex2}).

\textbf{These findings suggest that participants rarely generated entirely novel ideas; instead, they adapted and specialized existing examples to better match their own workflows.}

\subsection{RQ2: How does malleability change participants’ inbox experiences?}

Malleability fundamentally changed how participants experienced their inbox, shifting it from a fixed interface into a customizable environment that could be shaped to support their workflows. Rather than using a single predefined view, participants created tailored representations that surfaced information in more useful forms. In this shift, the inbox itself became less of a fixed interface and more of an underlying \textit{database} of information, while user-authored customizations became the primary interface through which that information was accessed.

Across participants, features reorganized existing information into more actionable formats. For example, P0 and P3 created separate feeds for newsletters, allowing them to view and filter articles independently of their inbox. P0 noted that while he could have accessed this information previously, the feature made “the process of trying to stay updated easier” by isolating relevant content. 

Participants also created features that transformed their inbox into task-oriented dashboards. P1 created an apartment tracking interface that surfaced listings in a structured view, enabling him to quickly review and act on new opportunities without searching through his inbox - having a dedicated page allowed him to “reach out immediately [to brokers] instead of having to find the listings in my inbox.” P2 described his job application tracker feature as replacing a manually maintained spreadsheet, automatically aggregating updates into a centralized view. P6’s feature surfaced deadlines and time estimates directly within the inbox, allowing him to “just see what I needed to do.” P6 specifically noted that his feature reminded him of multiple deadlines he had forgotten. In these cases, the inbox was no longer simply a collection of messages, but a system for tracking and acting on ongoing work.

In other cases, malleability enabled participants to separate and prioritize different types of work that were previously interleaved. P7 described feeling “overwhelmed” by the mixture of emails in her inbox, and created a feature that reorganized advisor communications by project. This allowed her to distinguish between different contexts (e.g., research vs. startup work) even when emails came from the same sender. 

Malleability also enabled delegation of low-priority or repetitive tasks. P4 created an auto-reply system to handle routine student emails, allowing them to offload what they described as “a pile of repetitive low-priority emails.”

Across these examples, malleability changed how information was structured, surfaced, and acted on. Participants moved from navigating a generic inbox interface to constructing personalized views that aligned with their priorities and workflows. \textbf{These findings suggest that malleability shifts the inbox from a fixed interface to a flexible data layer, where user-authored features become the primary interface.}

\subsection{RQ3: What risks emerge from participant-authored features, and how do participants reason about them?}

While participants were able to successfully create and use custom features, these features also introduced new forms of risk stemming from the fact that users were directly shaping system behavior.

One common risk was incorrect or incomplete system behavior, particularly in features that involved classification or inference. For example, P6’s deadline-surfacing feature occasionally surfaced emails that were not actually important or relevant deadlines. P6 could mark those as incorrect but expressed a desire for the system to learn from these corrections over time. This reflects an expectation that user-authored features may be imperfect, but should be \textit{adaptable} through lightweight feedback mechanisms.

Participants also raised concerns about missing important information due to filtering or restructuring. Several features involved isolating specific types of emails (e.g., apartment listings, newsletters, or events) into separate views, which improved usability but introduced the risk of hiding relevant content. Participants explicitly referenced worst-case scenarios such as missing a cheap apartment or an interview, suggesting that participants did consider the possibility of catastrophic failure. As a result, users implicitly expected mechanisms for auditing or recovering filtered information.

Another class of risk emerged from selective over-reliance on automated behavior. P7, who created an auto-reply feature for handling routine student emails, reported not verifying the generated responses. In this case, the system enabled delegation of low-priority work, but also introduced the possibility of unnoticed errors. Importantly, participants’ willingness to trust the system was highly dependent on task importance, suggesting that users naturally calibrate oversight based on perceived stakes.

Across these examples, participants did not expect their features to be perfectly correct or complete at creation. Instead, they reasoned about risk in terms of whether they could \textit{see}, \textit{correct}, and \textit{adapt} system behavior over time. This suggests that the primary challenge in malleable systems is not eliminating errors, but supporting users in managing and recovering from them, particularly as systems are used over longer periods and in higher-stakes contexts. Importantly, participants were largely accepting of these risks and did not report them as severe enough to stop using their features; however, this acceptance should be interpreted within the scope of the study.

\textbf{Risks in malleable systems arise not only from incorrect behavior, but from limited visibility, lack of adaptability, and the consequences of failure; users expect to mitigate these risks through feedback, oversight, and iterative refinement.}
\section{Discussion}

\subsection{From Frustration to Feature/Fix}

A central shift enabled by malleable systems is the ability to move directly \textit{from frustration to feature/fix}. In traditional software, moments of breakdown—such as difficulty locating information or responding efficiently—rarely translate into immediate change. Instead, they remain unresolved or require external tooling, technical expertise, or feature requests that may never be implemented.

In contrast, participants in our study were able to identify breakdowns in their inbox experience and rapidly construct features that addressed them. These features were not speculative; they emerged directly from situated use. Participants often described a vague sense that something was ``missing'' or ``hard to find,'' and used the system to externalize that feeling into a concrete interface change. This tight feedback loop—where problems are immediately actionable—suggests a new interaction paradigm in which users continuously refine their tools as they use them. This suggests a shift from systems that are used despite their limitations to systems that can be continuously reshaped in response to them.

This stands in contrast to prior work in end-user customization, which has found that users rarely customize systems due to high barriers or difficulty translating needs into implementations. By embedding customization directly into everyday workflows and lowering the cost of expression to natural language, malleable systems fundamentally alter this dynamic. \textbf{Malleable systems enable a direct loop from breakdown to repair, allowing users to translate frustration into features and fixes in real time.}

\subsection{Crafting Personal Information Environments}

Our findings suggest that malleable systems allow users to treat software not as a fixed interface, but as a workspace they can actively construct and refine. This mirrors how individuals organize physical workspaces, arranging tools and materials so that relevant resources are readily accessible in the forms most useful to them. In this framing, systems like Gmail are not inherently tied to their default interfaces; rather, they are underlying data stores presented through a particular view. Participants’ customizations demonstrate that the same underlying data can support multiple, coexisting interfaces tailored to different goals—from summarizing content to surfacing time-sensitive information.

This shift—from interface as fixed product to interface as user-authored layer—suggests that systems should enable users to curate their own information environments, shaping how data is surfaced, organized, and acted upon. \textbf{Malleable systems allow users to construct personalized information environments, treating underlying data as stable while reshaping the interface around their needs.}

\subsection{Malleability and the Design Process}

Malleable systems blur the boundary between \textit{use} and \textit{design}. Participants designed features in situ, while interacting with their inbox. This suggests a fundamentally different design process—one that is continuous, user-driven, and tightly coupled to real-world use.

Rather than requiring users to anticipate needs in advance, malleable systems support incremental, iterative design grounded in lived experience. Participants could test features immediately, observe their effects, and refine them over time. In this way, the system enables a form of \textit{end-user design-in-use}, where the act of using a tool becomes inseparable from shaping it. This reduces the gap between identifying a problem and implementing a solution, enabling users to act on needs as they arise rather than deferring them.

\paragraph{Implications for Design.}
These findings suggest several implications for the design of malleable systems. First, systems should support \textit{low-friction expression of intent}, enabling users to translate vague or emerging frustrations into actionable system changes. Second, systems should enable \textit{immediate execution and feedback}, so that users can evaluate features within the context of their ongoing workflows. Third, malleable systems should support \textit{incremental refinement}, allowing users to iteratively adapt features rather than requiring fully-formed designs upfront. Fourth, systems should support \textit{safe exploration}, allowing users to experiment with new functionality without fear of breaking existing workflows or losing important information.

More broadly, malleable systems shift design from static interfaces to systems that enable interface construction. In this paradigm, designers are no longer solely responsible for anticipating user needs; instead, they build infrastructures that allow users to author and evolve their own tools over time. This reframes human-centered design as an ongoing, participatory process embedded within everyday interaction. \textbf{Malleable systems transform design from a one-time activity into a continuous, in-use process where users iteratively construct and refine their own tools.}

\subsection{Perceived Control and Engagement}

Participants consistently reported feeling a greater sense of control over their inbox. This was not simply due to automation, but to the ability to define how their inbox behaved. By creating features that aligned with their priorities and workflows, participants transformed their relationship to the system—from passive recipients of a fixed interface to active shapers of their experience.

Notably, features could take multiple forms: some were embedded directly within the inbox, while others manifested as entirely new interfaces derived from it. This flexibility allowed participants to choose the form of intervention that best matched their needs. 

Overall, participants responded positively to these capabilities, suggesting that malleability may increase not only efficiency but also engagement and satisfaction with everyday tools. \textbf{By enabling users to shape system behavior directly, malleable systems increase perceived control and foster deeper engagement with everyday tools.}

\subsection{Limitations}

This work is based on a short-term design probe and should be interpreted within that scope. While participants were able to successfully create and use features over multiple days, several aspects of the study shape how these findings should be understood.

First, the study duration was relatively brief. While this allowed us to capture initial feature creation and early use, participants may not have experienced longer-term dynamics such as evolving needs, sustained maintenance, or changes in feature usefulness over time.

Second, the system was implemented as a research prototype. Although it supported feature creation and iteration, some interactions involved guided assistance, which may have influenced how participants approached refinement. Future systems with more fully self-contained tooling may better reflect how users independently iterate on their own customizations.

Third, participants primarily focused on everyday, relatively low-stakes use cases. As a result, the risks observed in this study may not fully capture challenges that could arise in higher-stakes or more critical contexts. While participants were generally comfortable using their features despite imperfections, this tolerance may differ in settings where errors carry greater consequences.

Finally, as a design probe, this study is intended to explore how users engage with malleable systems rather than to evaluate system performance or long-term adoption. Accordingly, our findings highlight emerging interaction patterns and behaviors, but do not yet speak to system robustness or sustained real-world deployment.

\subsection{Future Work}
Future work should explore malleable systems in more mature and longitudinal contexts. A key next step is to deploy a more stable, fully user-driven system over longer periods of time to understand how features evolve, accumulate, and interact. This would enable investigation into questions such as how users manage a growing set of custom features, how needs change over time, and what mechanisms support stability and trust.

Additionally, this work focused on email as a domain, but the underlying principles may extend to other productivity tools. Future research could examine malleability in contexts such as calendars, document workflows, project management systems, or personal knowledge bases, where users similarly interact with large, structured datasets through fixed interfaces.

Finally, further work is needed to better understand and mitigate risks. This includes developing mechanisms for feature validation, transparency, and error handling, as well as exploring how users reason about and debug their own customizations. 

Future work should examine how malleable systems scale over time, across domains, and under higher-stakes conditions. Ultimately, realizing their full potential will require rethinking how interfaces are designed, deployed, and evolved over time.
\section{Conclusion}
We presented a design probe exploring conversational customization within a high-frequency productivity tool. By enabling users to modify system behavior through natural language, our system made customization an in-situ, ongoing activity rather than a one-time configuration step. Across participants, we observed that feature creation was grounded in everyday friction, shaped by existing patterns, and refined through real use. These customizations did not introduce new information, but reshaped how it was structured, surfaced, and acted upon.

Our findings suggest that malleability changes not only what users can do with a system, but how they relate to it. Participants moved from adapting to a fixed interface toward actively constructing representations of their own information environment, while reasoning about trade-offs in correctness, visibility, and risk. In this way, customization becomes part of interaction itself, rather than an auxiliary capability.

Taken together, this work points toward a broader shift in how productivity systems are designed: from static interfaces defined in advance, to malleable environments that are continuously shaped through use. As systems become increasingly capable, the central challenge is no longer what they can do, but how effectively users can shape them to meet their evolving needs.

\section{Citations}

\bibliographystyle{ACM-Reference-Format}
\bibliography{references}

\appendix
\onecolumn

\clearpage
\section{User Interaction Model}
\label{usermodel}
\begin{figure}[H]
\centering
\includegraphics[width=\textwidth]{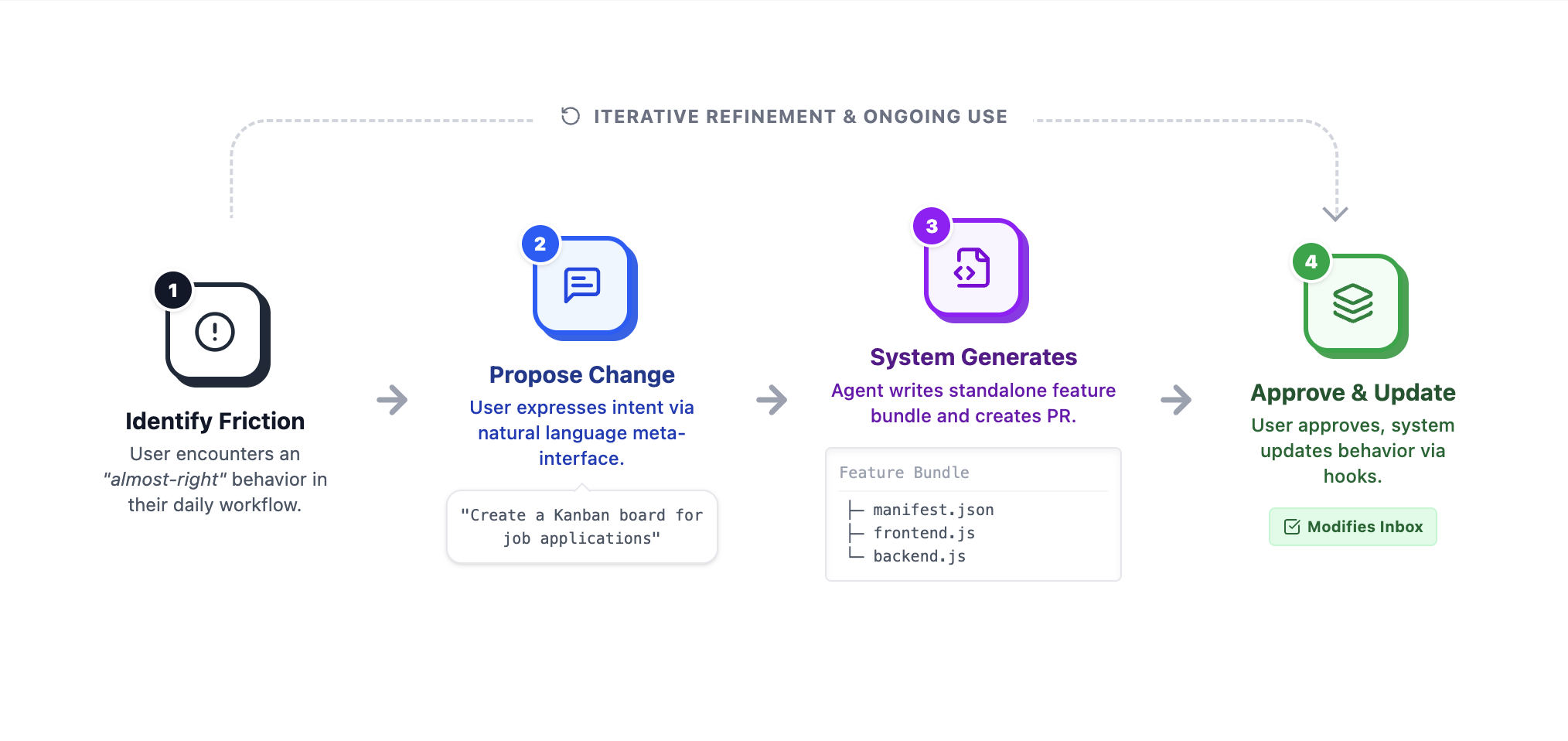}
\caption{
The conversational customization loop. Users identify friction in their daily workflow, propose a change through the natural language meta-interface, and the system generates a standalone feature bundle. After user approval, the feature is deployed and modifies inbox behavior. This cycle repeats as users refine features through ongoing use.
}
\label{fig:interaction-loop}
\Description{
A four-step diagram showing the conversational customization loop: (1) Identify Friction, where the user encounters almost-right behavior, (2) Propose Change, where the user expresses intent via natural language, (3) System Generates, where the agent writes a standalone feature bundle containing manifest.json, frontend.js, and backend.js, and (4) Approve and Update, where the user approves and the system updates behavior. A dashed arrow loops back from step 4 to step 1, labeled Iterative Refinement and Ongoing Use.
}
\end{figure}
\clearpage

\clearpage
\section{System Architecture}
\label{arch}
\begin{figure}[H]
\centering
\includegraphics[width=\textwidth]{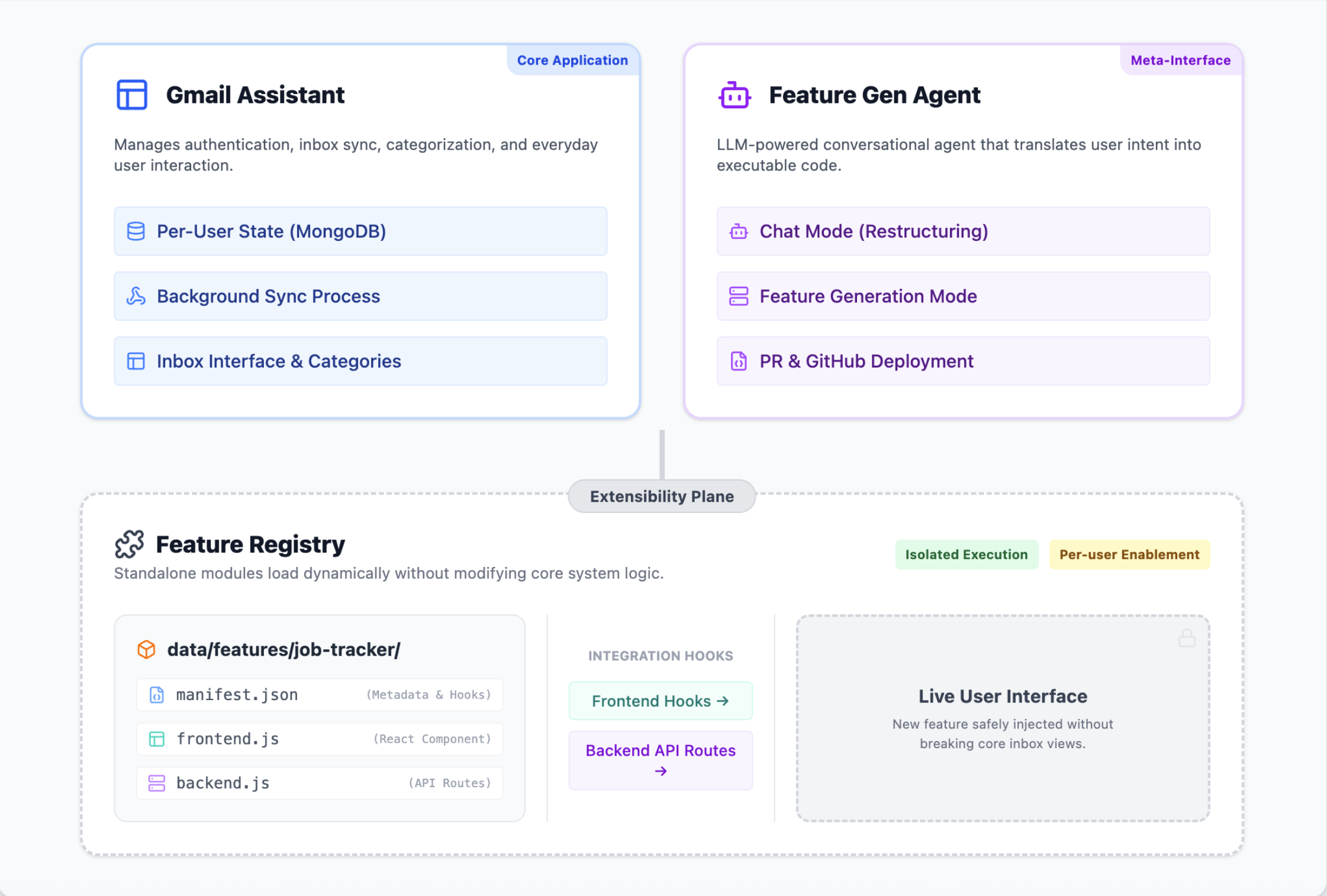}
\caption{
System architecture. The Gmail Assistant (left) manages authentication, inbox sync, categorization, and everyday interaction. The Feature Gen Agent (right) translates user intent into executable code through chat-based restructuring and feature generation modes, with deployment via GitHub. Both connect to an extensibility plane where standalone feature modules load dynamically through frontend hooks and backend API routes without modifying core system logic.
}
\label{fig:system-architecture}
\Description{
A system architecture diagram showing two top-level components: the Gmail Assistant managing per-user state, background sync, and the inbox interface, and the Feature Gen Agent supporting chat mode, feature generation mode, and PR-based deployment. Below both is an extensibility plane containing a feature registry where standalone modules with manifest.json, frontend.js, and backend.js integrate into the live user interface through frontend hooks and backend API routes.
}
\end{figure}

\clearpage
\section{Feature Registry}
\label{app:registry}

\begin{figure}[H]
\centering
\includegraphics[width=\textwidth]{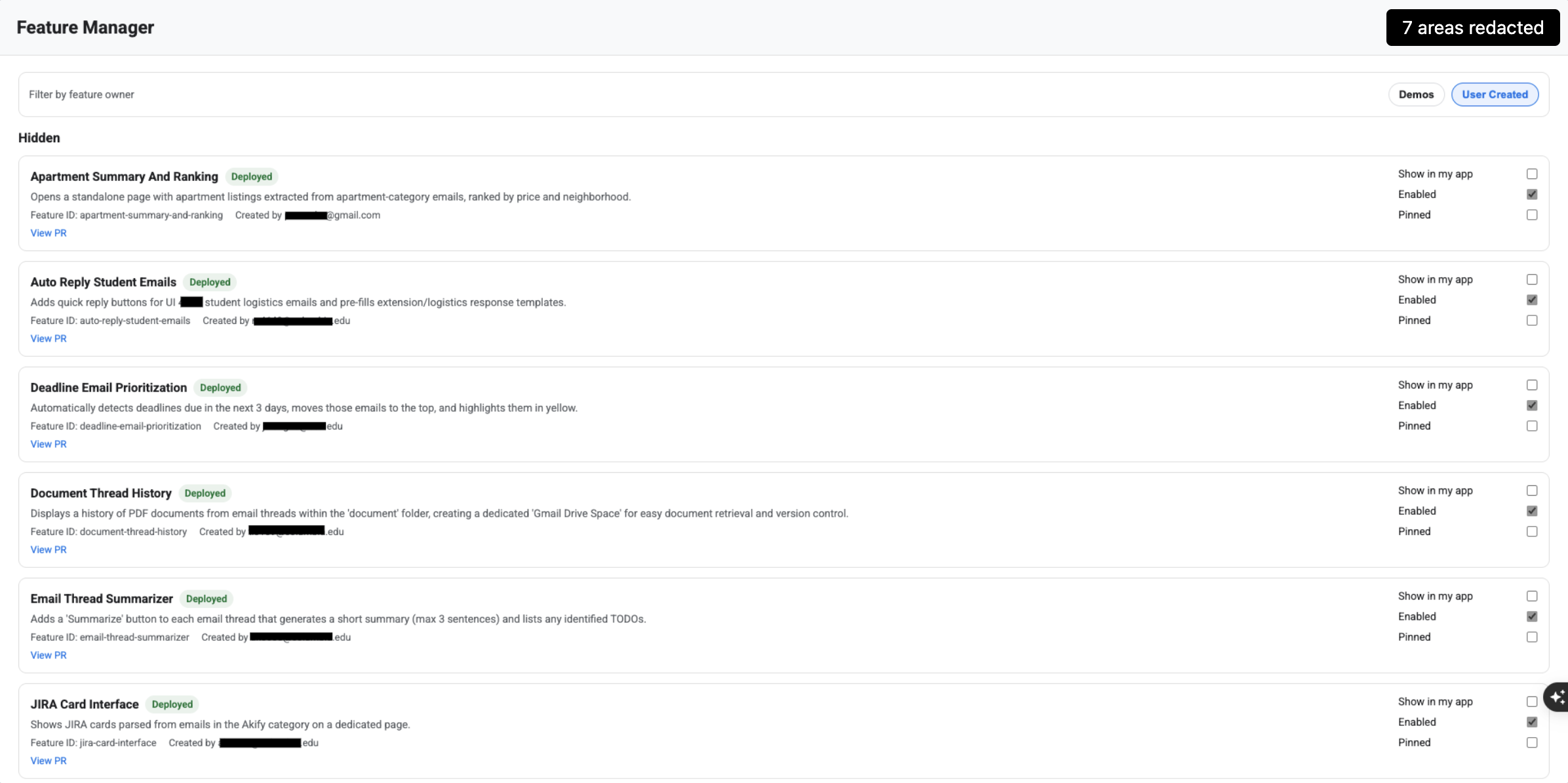}
\caption{
Feature manager interface. Deployed features are listed with descriptions, ownership metadata, and controls for visibility, enablement, and pinning. This registry provides a persistent, inspectable layer for managing user-authored functionality after deployment. Features are sorted into as being created as demos or by users of the system.
}
\label{fig:feature-registry}
\Description{
Screenshot of the feature manager interface. The page lists several deployed features, including Apartment Summary and Ranking, Auto Reply Student Emails, Deadline Email Prioritization, Document Thread History, Email Thread Summarizer, and JIRA Card Interface. Each entry includes a description, feature ID, creator, and checkboxes to control whether the feature is shown in the app, enabled, or pinned.
}
\end{figure}

\clearpage
\section{Example Feature \#1: Category Creation}
\label{ex1}
\begin{figure}[H]
    \centering
    \includegraphics[width=\linewidth]{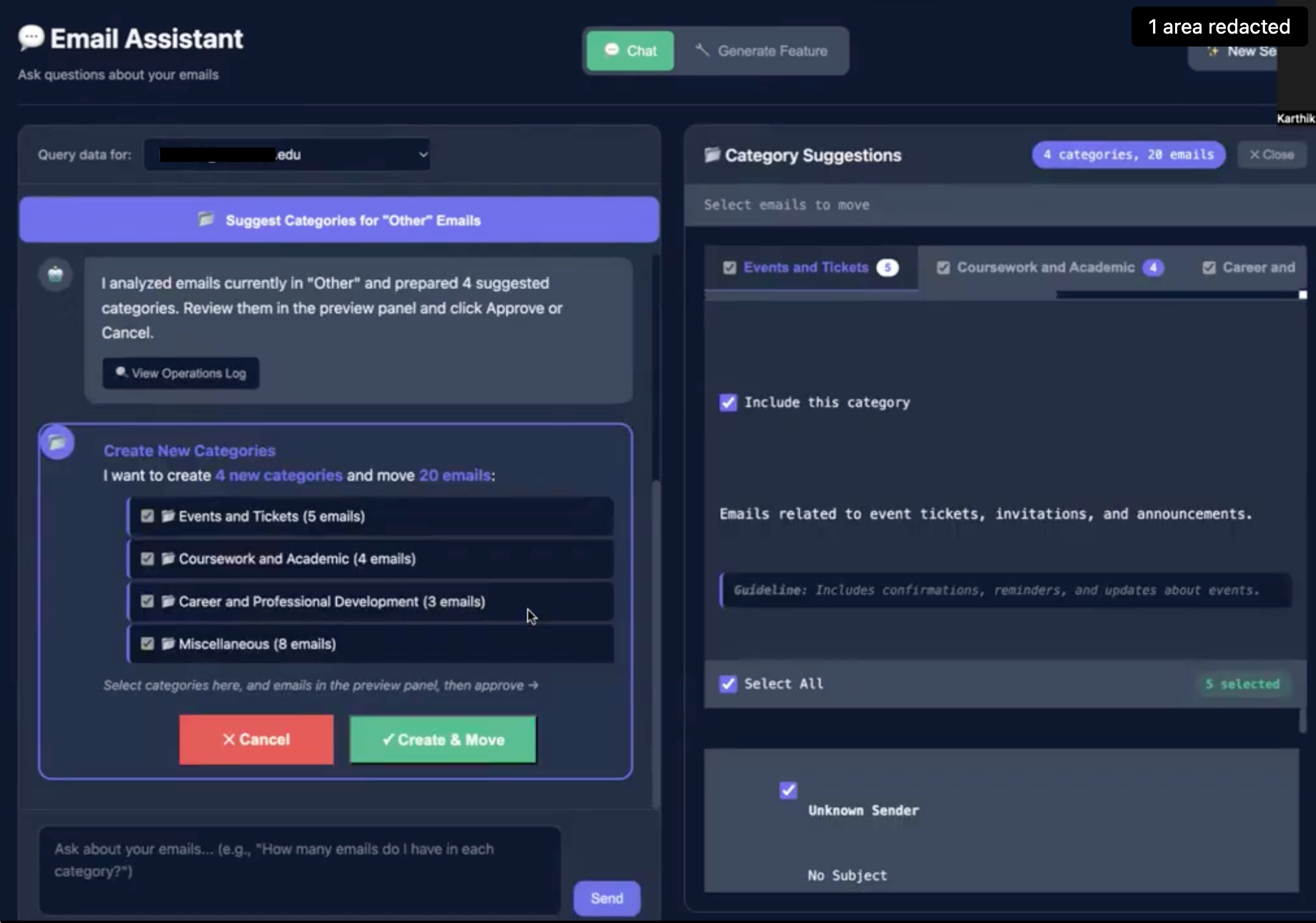}
    \caption{\textbf{Conversational restructuring interface for inbox organization.} 
    In chat mode, the agent analyzes emails in ambiguous categories (e.g., ``Other'') and proposes a set of new categories along with candidate email assignments. The interface presents these suggestions for user review, allowing categories to be selectively approved before emails are reorganized. This interaction enables users to restructure their inbox through natural language while maintaining control via confirmation-based workflows.}
    \Description{A chat-based interface where the system proposes new email categories and shows candidate email assignments, allowing the user to review and approve reorganization actions.}
    \label{figures:ex1}
\end{figure}

FOR ALL EXAMPLE FEATURES, VIDEOS WILL BE INCLUDED IN AN UNANONYMIZED SUBMISSION.

\clearpage
\section{Example Feature \#2: Quick Reply}
\label{ex2}
\begin{figure}[H]
    \centering
    \includegraphics[width=\linewidth]{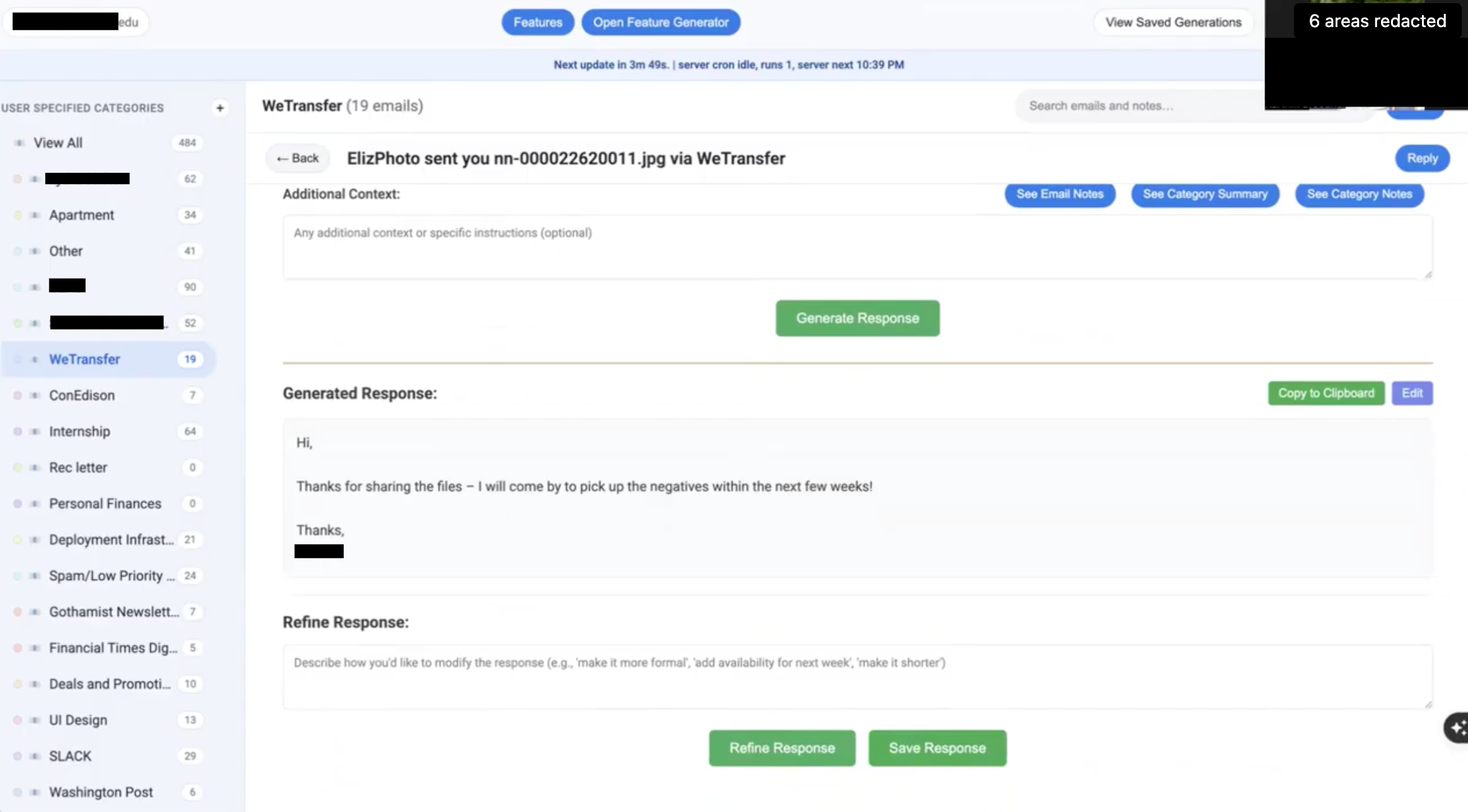}
    \caption{\textbf{Category-scoped quick reply interface for email response generation.} 
    Within a specific category (e.g., WeTransfer), the system prefills the response window with a pre-defined user template.}
    \Description{An email interface showing a generated reply draft for a message within a specific category, with options to add context, refine the response, and copy or save the result.}
    \label{figures:ex2}
\end{figure}

\clearpage
\section{Example Feature \#3: ToDos on Emails}
\label{ex3}
\begin{figure}[H]
    \centering
    \includegraphics[width=\linewidth]{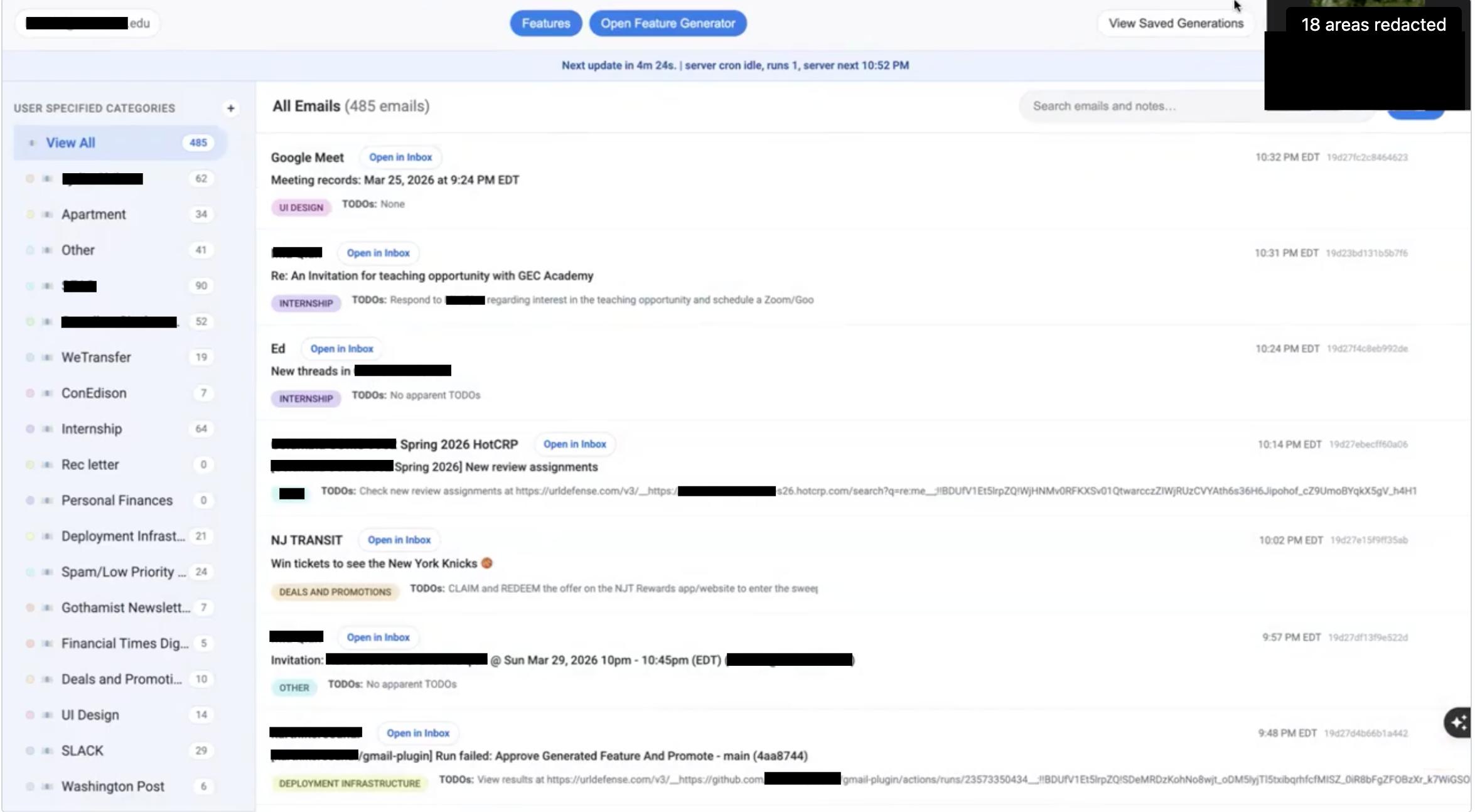}
    \caption{\textbf{Inline TODO extraction and augmentation within the inbox.} 
    The system analyzes email content to identify actionable items and surfaces them directly alongside each message as ``TODOs.'' These extracted tasks (e.g., responding, scheduling, reviewing links) are displayed inline within the inbox, enabling users to quickly understand required actions without opening individual emails. By augmenting messages with structured action summaries, the feature transforms the inbox from a passive communication log into an actionable task management interface.}
    \Description{An email inbox interface where messages are annotated with extracted TODOs, summarizing actions such as responding, scheduling, or reviewing content directly within the message list.}
    \label{figures:ex3}
\end{figure}

\clearpage
\section{Example Feature \#4: Category Sub-Filter}
\label{ex4}
\begin{figure}[H]
    \centering
    \includegraphics[width=\linewidth]{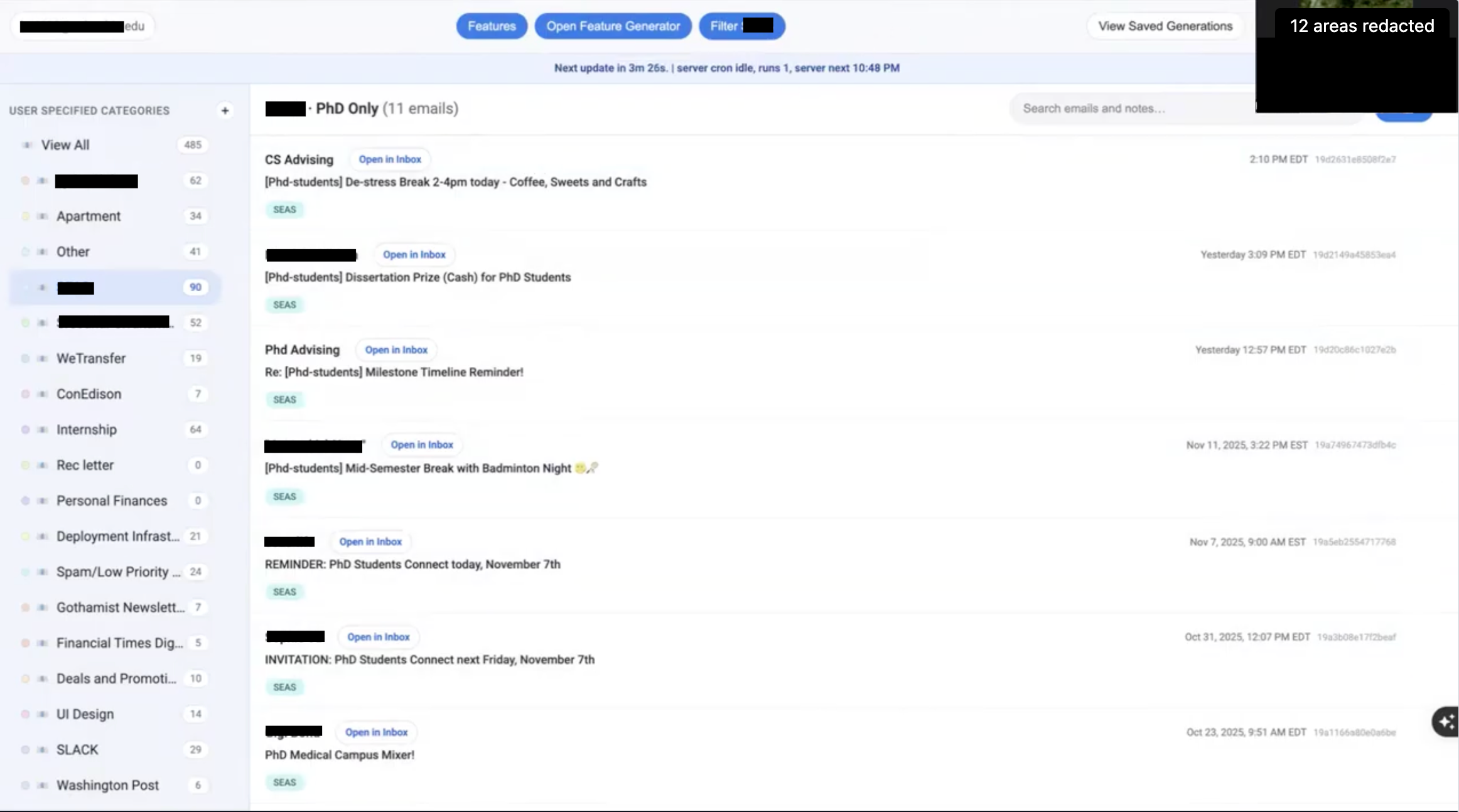}
    \caption{\textbf{Category refinement for surfacing PhD-specific emails.} 
    The system enables users to define more granular filters within broader categories (e.g., university-related emails), isolating messages relevant to specific subgroups such as PhD students. The resulting ``PhD Only'' category surfaces targeted communications (e.g., advising updates, events, and opportunities) without requiring manual search through the broader category. This feature supports personalized restructuring of the inbox to reflect user-specific contexts and roles.}
    \Description{An email inbox interface showing a filtered category labeled ``PhD Only,'' containing emails relevant to PhD students extracted from a broader set of university-related messages.}
    \label{figures:ex4}
\end{figure}

\clearpage
\section{Example Feature \#5: Surface Emails Requiring Response}
\label{ex5}
\begin{figure}[H]
    \centering
    \includegraphics[width=\linewidth]{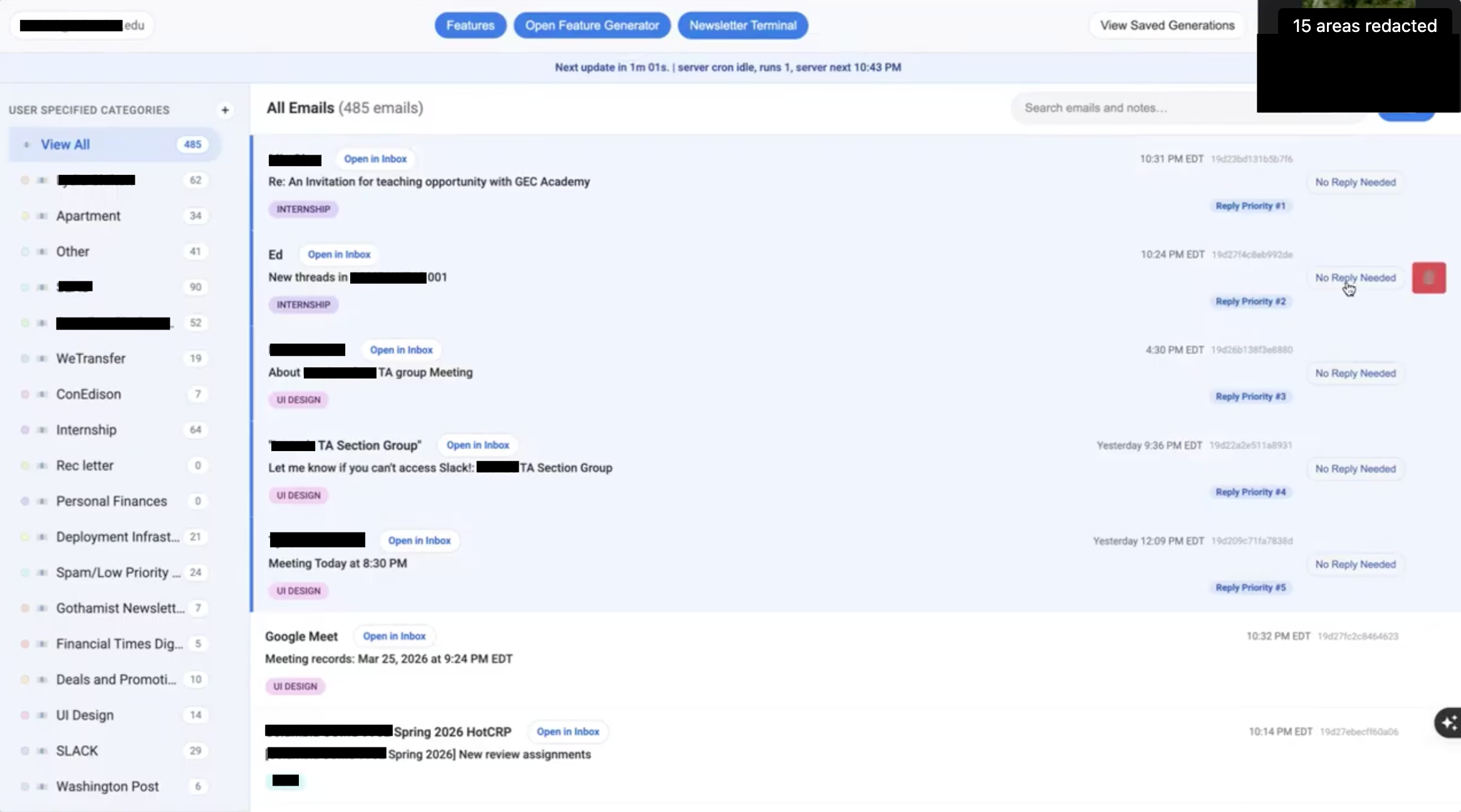}
    \caption{\textbf{Reply prioritization interface for surfacing emails that require responses.} 
    The system analyzes incoming emails to determine which messages require a reply and ranks them by priority, surfacing these emails at the top of the inbox. Messages are annotated with indicators such as ``Reply Priority'' and ``No Reply Needed,'' allowing users to quickly focus on actionable communication while deprioritizing informational emails. This feature introduces an attention management layer that helps users triage their inbox based on response requirements.}
    \Description{An email inbox interface where certain messages are surfaced at the top and labeled with reply priority indicators, while others are marked as not requiring a response.}
    \label{figures:ex5}
\end{figure}

\clearpage
\section{Example Feature \#6: Financial Times Digest}
\label{ex6}
\begin{figure}[H]
    \centering
    \includegraphics[width=\linewidth]{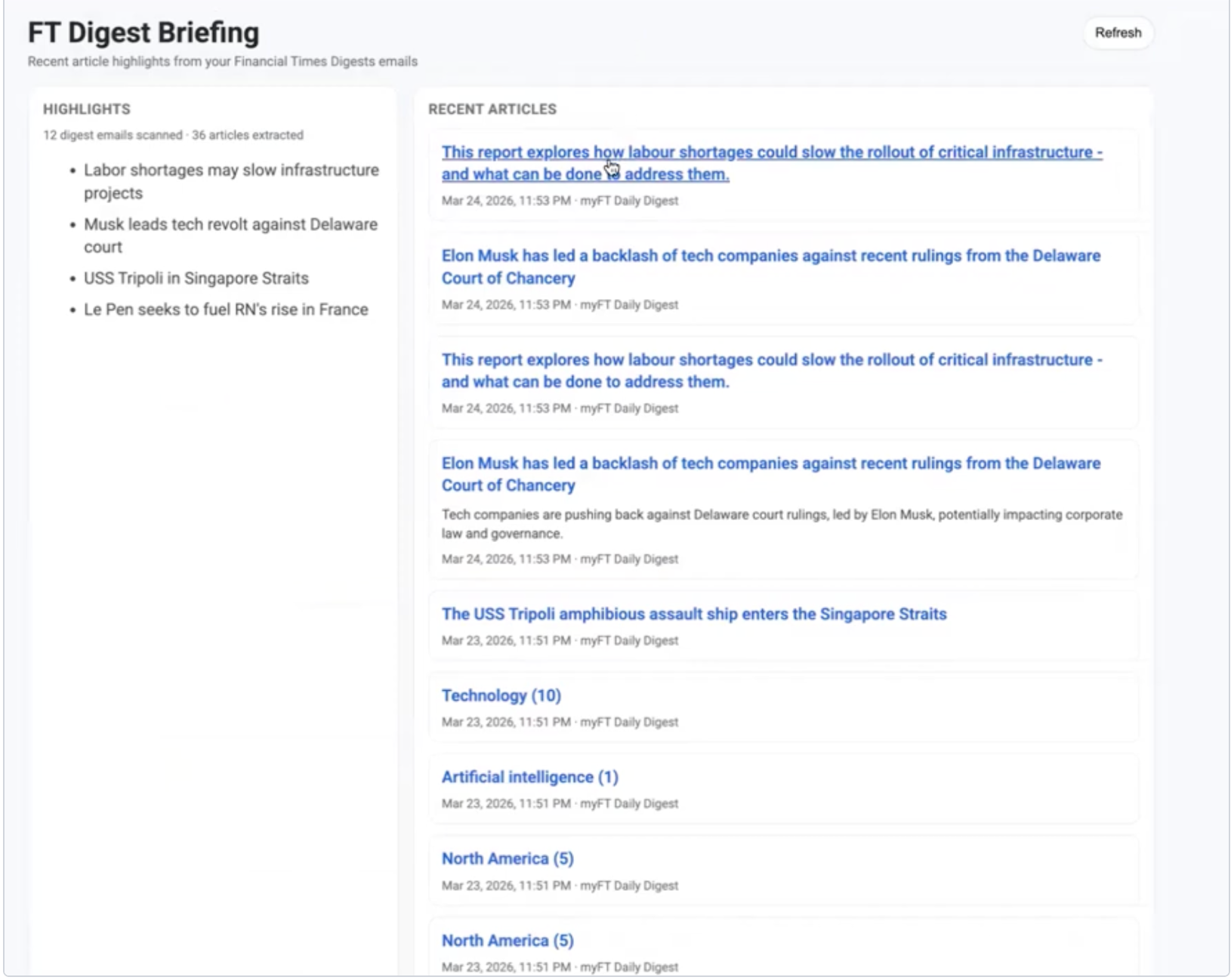}
    \caption{\textbf{Derived briefing interface for Financial Times newsletters.} 
    The system extracts articles from Financial Times digest emails and aggregates them into a structured briefing view, including highlighted summaries and a list of recent articles. Rather than requiring users to open individual newsletters, the interface synthesizes content across multiple emails, enabling efficient browsing of key stories and topics. This feature demonstrates how email content can be transformed into a dedicated reading interface tailored to user interests.}
    \Description{A dashboard interface displaying extracted articles from Financial Times newsletters, including a highlights panel and a list of recent articles with timestamps and categories.}
    \label{figures:ex6}
\end{figure}

\clearpage
\section{P0 | Design Probe}
\label{designprobe:p0}


\begin{figure}[H]
\centering
\includegraphics[width=\linewidth]{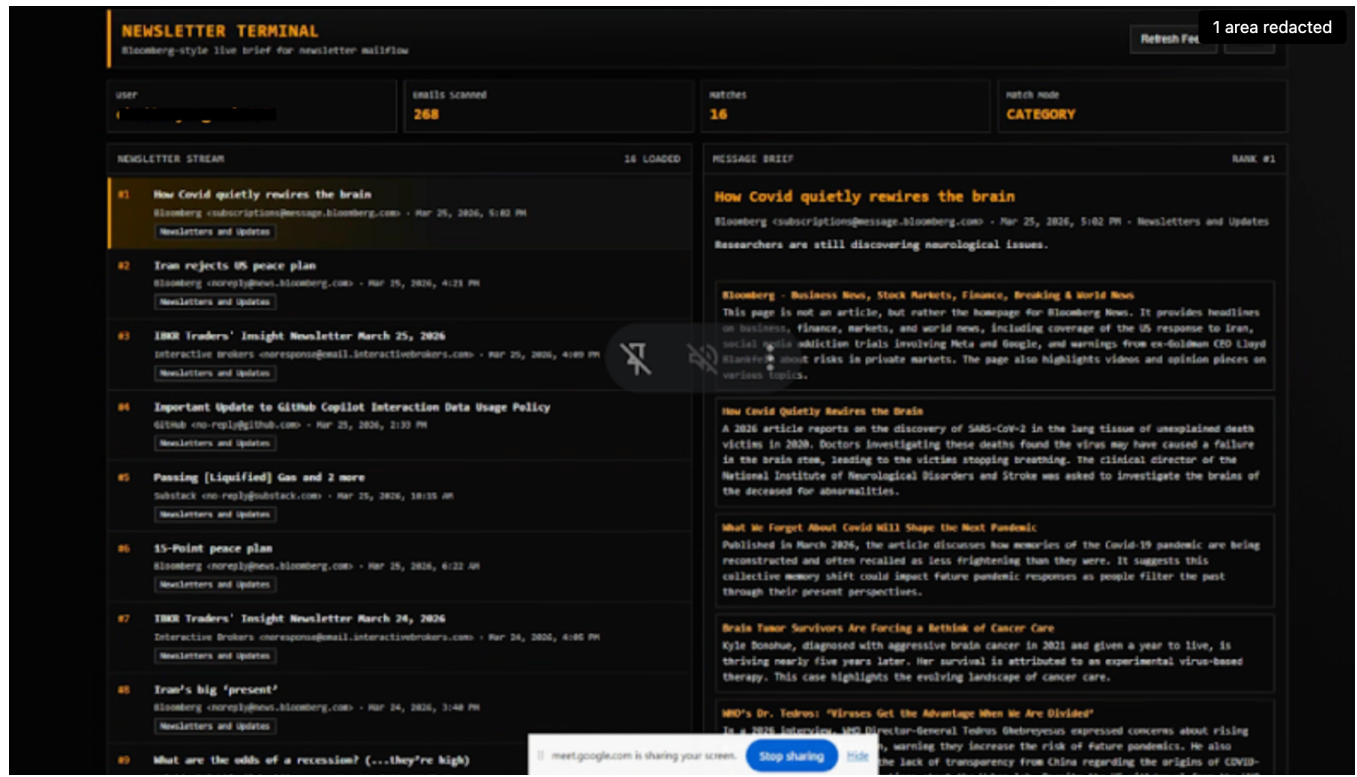}
\caption{
P0’s feature after Day 1: a “Bloomberg terminal-style” newsletter interface that aggregates emails into a standalone view with extracted articles and summaries. The interface separates content into a list of articles and a detailed reading pane, reflecting the participant’s goal of quickly scanning highlights from financial newsletters without opening individual emails.
}
\label{fig:day1p0appendix}
\end{figure}



\begin{figure}[t]
\centering
\includegraphics[width=\linewidth]{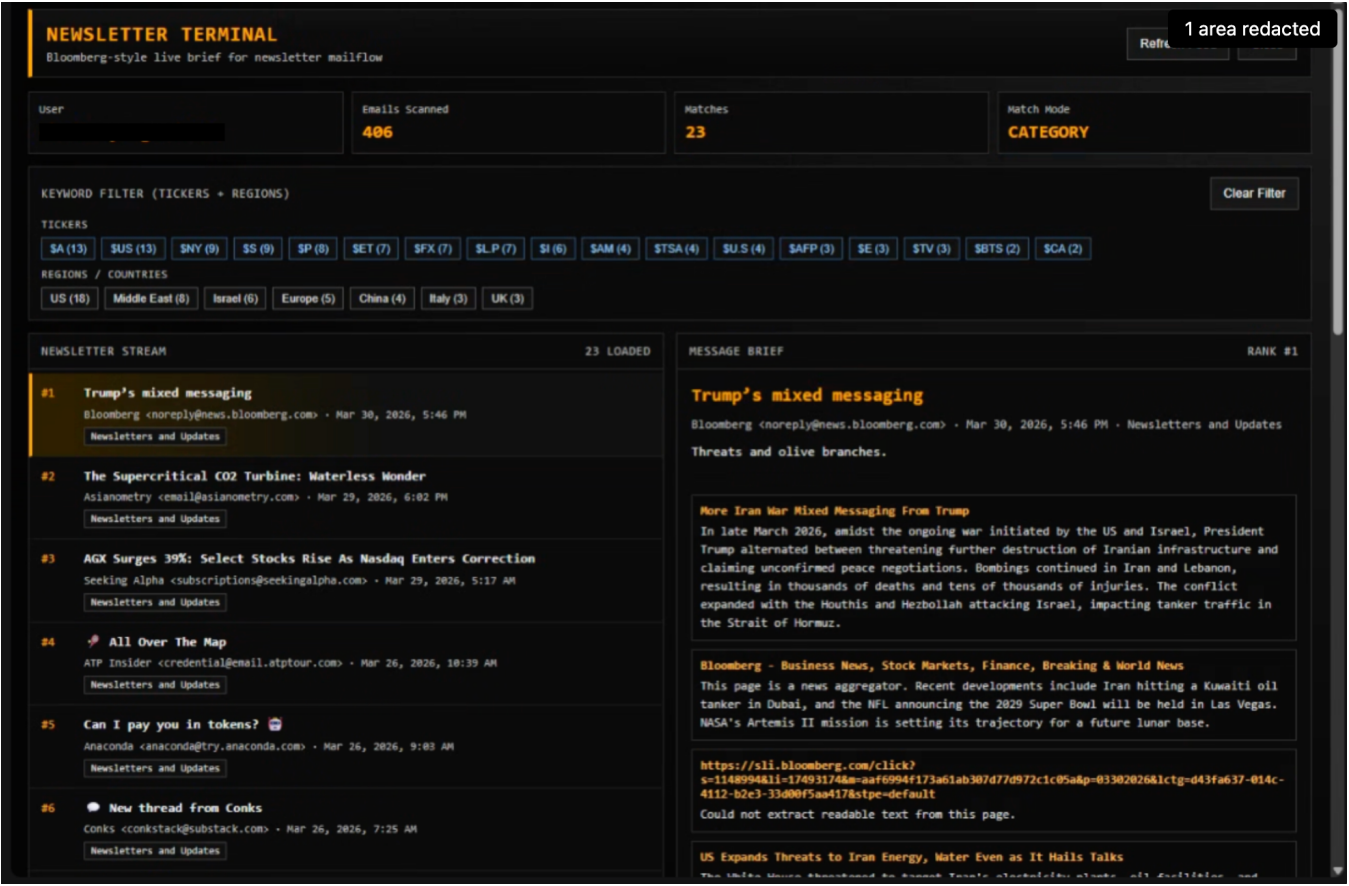}
\caption{
P0’s feature after Day 2: the newsletter interface extended with extracted financial tickers and geographic regions, enabling keyword-based filtering and more targeted exploration of content. It reflects refinement after real use, shifting the interface from general summarization toward domain-specific analysis.
}
\label{fig:day2p0appendix}
\end{figure}

\clearpage
\section{P1 | Design Probe}
\label{designprobe:p1}


\begin{figure}[H]
    \centering
    \includegraphics[width=\columnwidth]{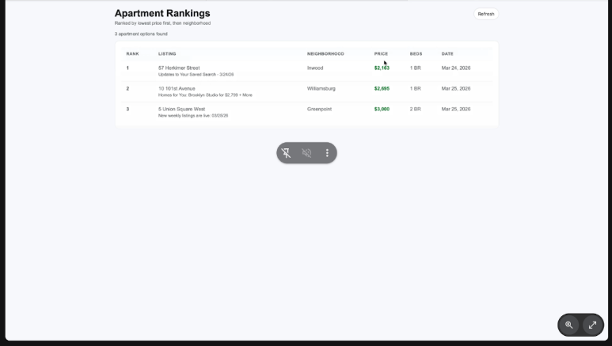}
    \caption{Participant P1’s Day 1 feature: an apartment listings ranking interface generated from email data. The interface extracts listings from incoming emails and presents them in a structured table with attributes such as neighborhood, price, number of bedrooms, and date, enabling quick comparison across options.}
    \Description{A web interface titled “Apartment Rankings” displaying a table of apartment listings with columns for rank, listing, neighborhood, price, number of bedrooms, and date.}
    \label{fig:day1p1}
\end{figure}



\begin{figure}[t]
    \centering
    \includegraphics[width=\columnwidth]{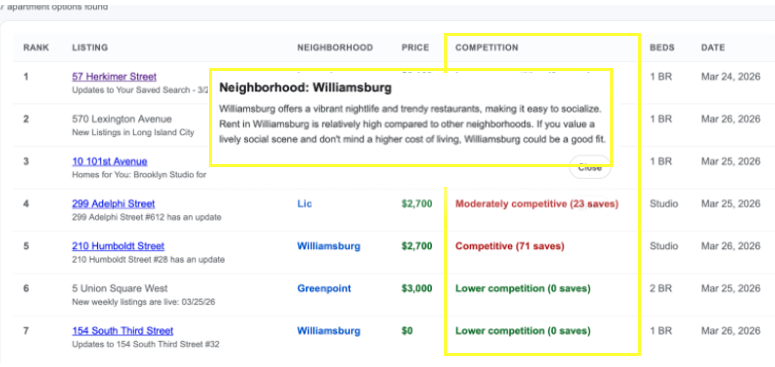}
    \caption{Participant P1’s Day 2 iteration of the apartment rankings feature. The system augments listings with inferred competition signals (e.g., number of saves) and labels apartments by competitiveness, while also supporting interactive neighborhood summaries that provide concise and tailored pros and cons. These additions are outlined in yellow on the UI.}
    \Description{An updated apartment rankings interface showing listings with additional fields such as competition level based on number of saves, and an expanded view displaying a short textual summary of a neighborhood’s characteristics.}
    \label{fig:day2p1}
\end{figure}

\clearpage
\section{P2 | Design Probe}
\label{designprobe:p2}

\begin{figure}[H]
    \centering
    \includegraphics[width=\linewidth]{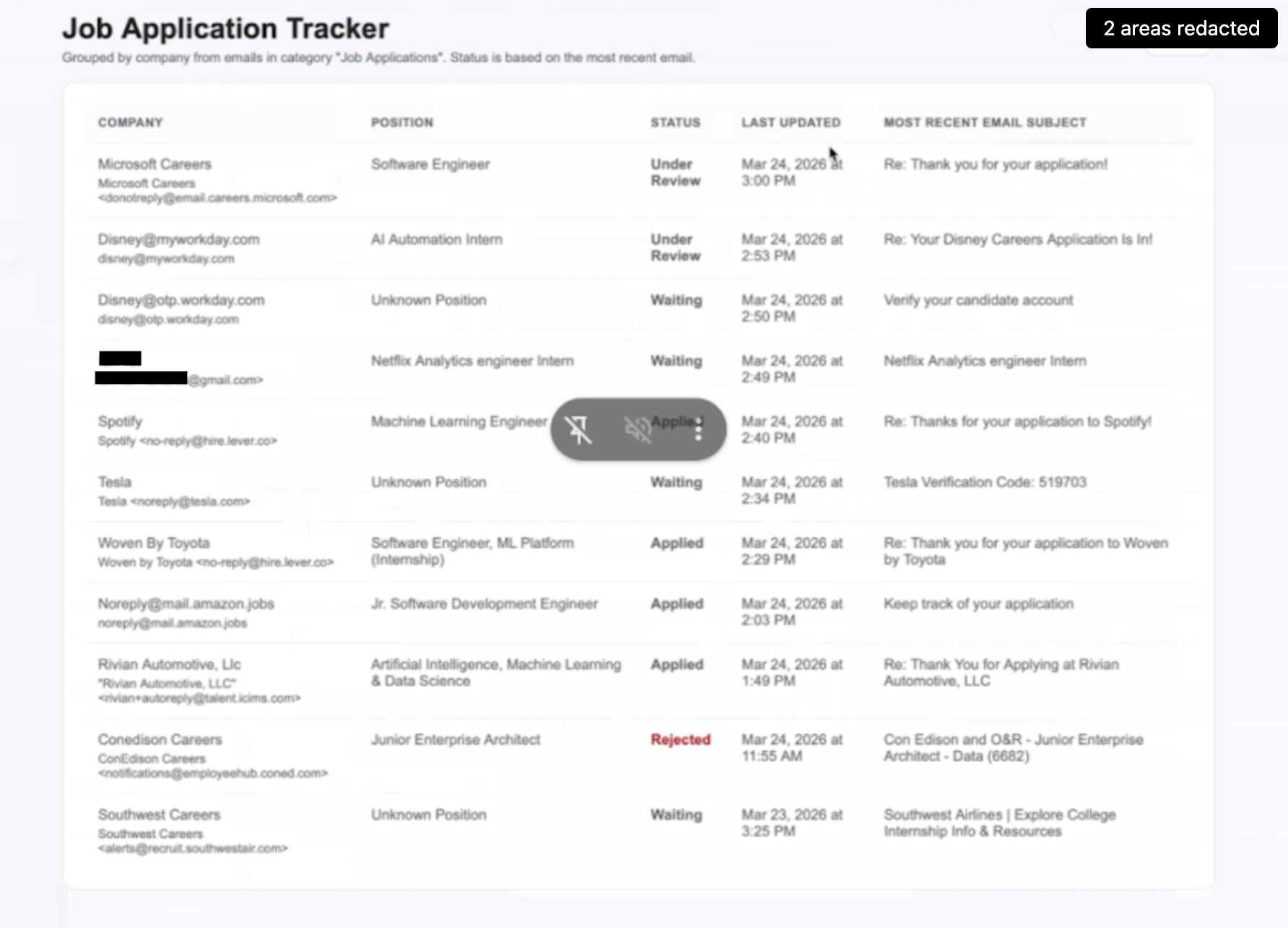}
    \caption{\textbf{Participant P2’s Day 1 job application tracker generated from their inbox.} 
    The system aggregates emails in the ``Job Applications'' category and groups them by company, extracting structured fields including role, status (e.g., applied, under review), last updated timestamp, and most recent email subject. This initial version reflects a transformation from an unstructured inbox into a centralized tracking interface, allowing P2 to monitor multiple applications without manually searching across threads.}
    \Description{A job application tracker interface listing companies, roles, statuses, timestamps, and recent email subjects, organized into a table derived from inbox emails.}
    \label{figures:day1p2}
\end{figure}

\begin{figure}[t]
    \centering
    \includegraphics[width=\linewidth]{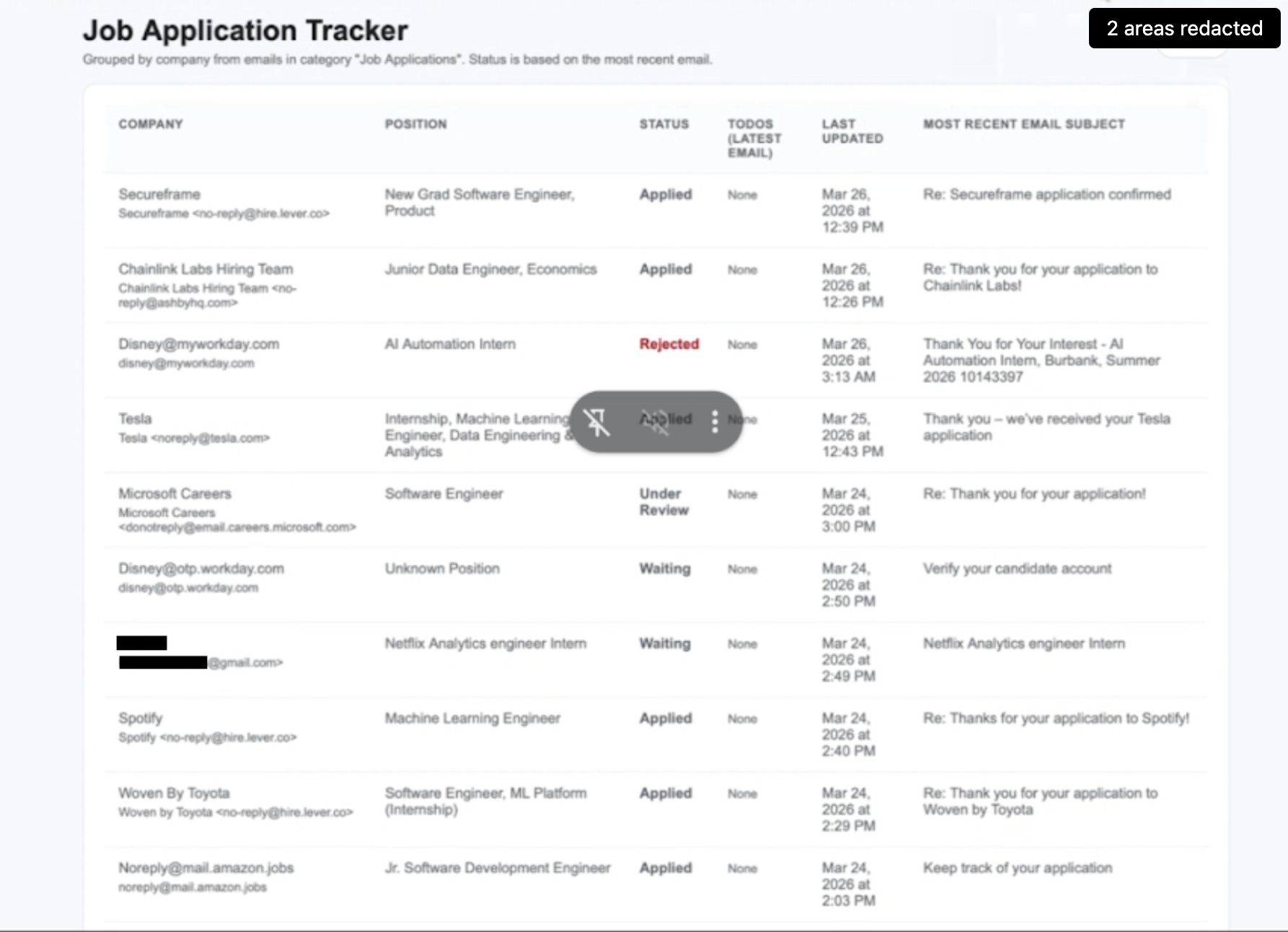}
    \caption{\textbf{Participant P2’s Day 2 iteration of the job application tracker introducing task awareness.} 
    In response to use, the feature is extended with a ``To-Dos'' column that surfaces actionable items inferred from the most recent email in each thread. These include reminders to follow up, complete application steps, or verify accounts. This modification shifts the interface from a passive summary of application states to an active workflow tool that highlights what actions are required next.}
    \Description{An updated job application tracker interface with an additional column displaying TODOs derived from recent emails, including follow-up reminders and pending actions.}
    \label{figures:day2p2}
\end{figure}

\begin{figure}[t]
    \centering
    \includegraphics[width=\linewidth]{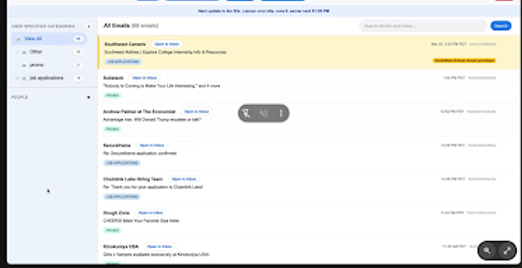}
    \caption{\textbf{Further iteration of P2’s feature introducing inbox-level prioritization behavior.} 
    Beyond the tracker interface, the system begins to intervene directly in the inbox by identifying application-related emails that have not received a response within a specified time window. These emails are surfaced and visually highlighted at the top of the inbox, along with suggestions to follow up. This extends the feature from a separate tracking view into modifying the user’s primary email workflow.}
    \Description{An email inbox interface where a job-related email is surfaced and highlighted at the top, indicating a suggested follow-up based on time since last response.}
    \label{figures:day2p2_1}
\end{figure}

\clearpage
\section{P3 | Design Probe}
\label{p3}
\begin{figure}[H]
    \centering
    \includegraphics[width=\linewidth]{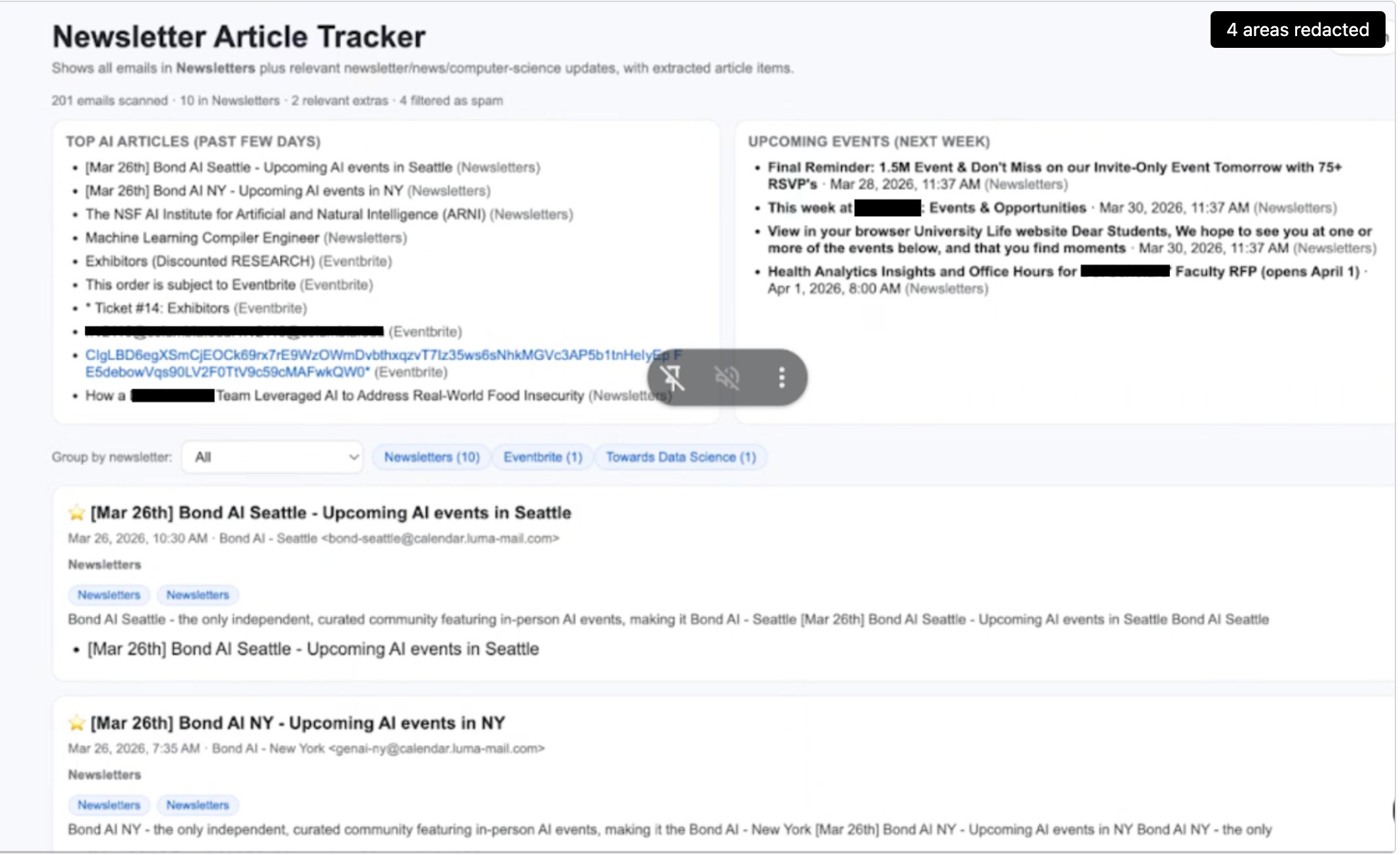}
    \caption{\textbf{Participant P3’s Day 1 newsletter article tracker derived from inbox emails.} 
    The system aggregates emails from the ``Newsletters'' category and extracts structured content, including top AI-related articles from recent emails and upcoming events mentioned across newsletters. Rather than presenting emails individually, the interface synthesizes content across sources into a dashboard view, allowing P3 to quickly scan relevant articles and events without opening each message.}
    \Description{A dashboard-style interface showing extracted newsletter content, including a list of top AI articles and a separate panel of upcoming events aggregated from multiple emails.}
    \label{figures:day1p3}
\end{figure}

\begin{figure}[t]
    \centering
    \includegraphics[width=\linewidth]{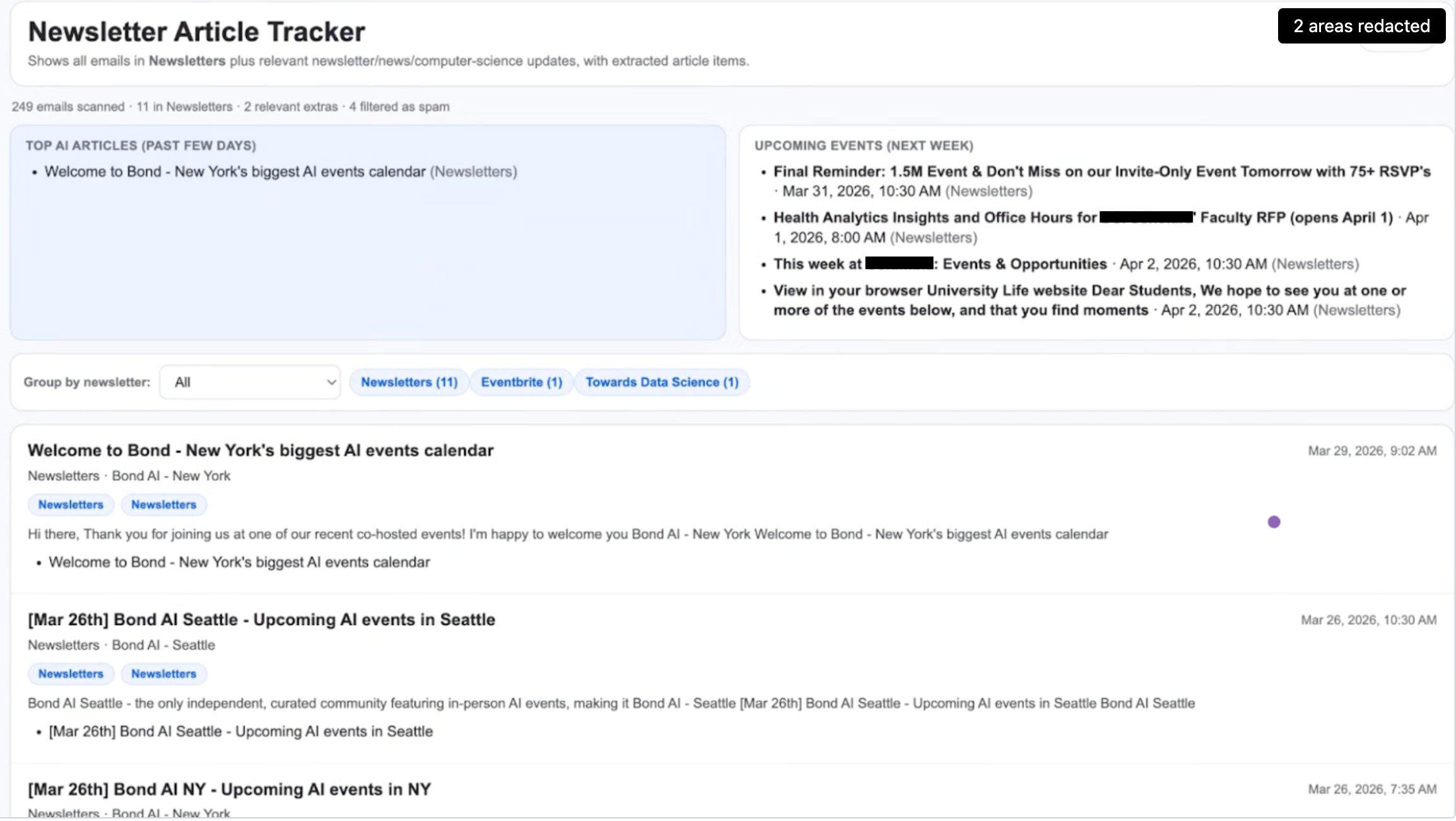}
    \caption{\textbf{Participant P3’s Day 2 iteration aligning the tracker with inbox conventions.} 
    Following initial use, the interface is updated to more closely match the formatting and structure of the native inbox, including visual styling and layout changes. While the underlying functionality remains unchanged, this iteration reflects a preference for consistency with existing email workflows, making the synthesized content easier to interpret within the user’s familiar interaction paradigm.}
    \Description{An updated newsletter dashboard interface with styling and layout adjusted to resemble a standard email inbox, while still displaying extracted articles and upcoming events.}
    \label{figures:day2p3}
\end{figure}

\clearpage
\section{P4 | Design Probe}
\begin{figure}[H]
    \centering
    \includegraphics[width=\linewidth]{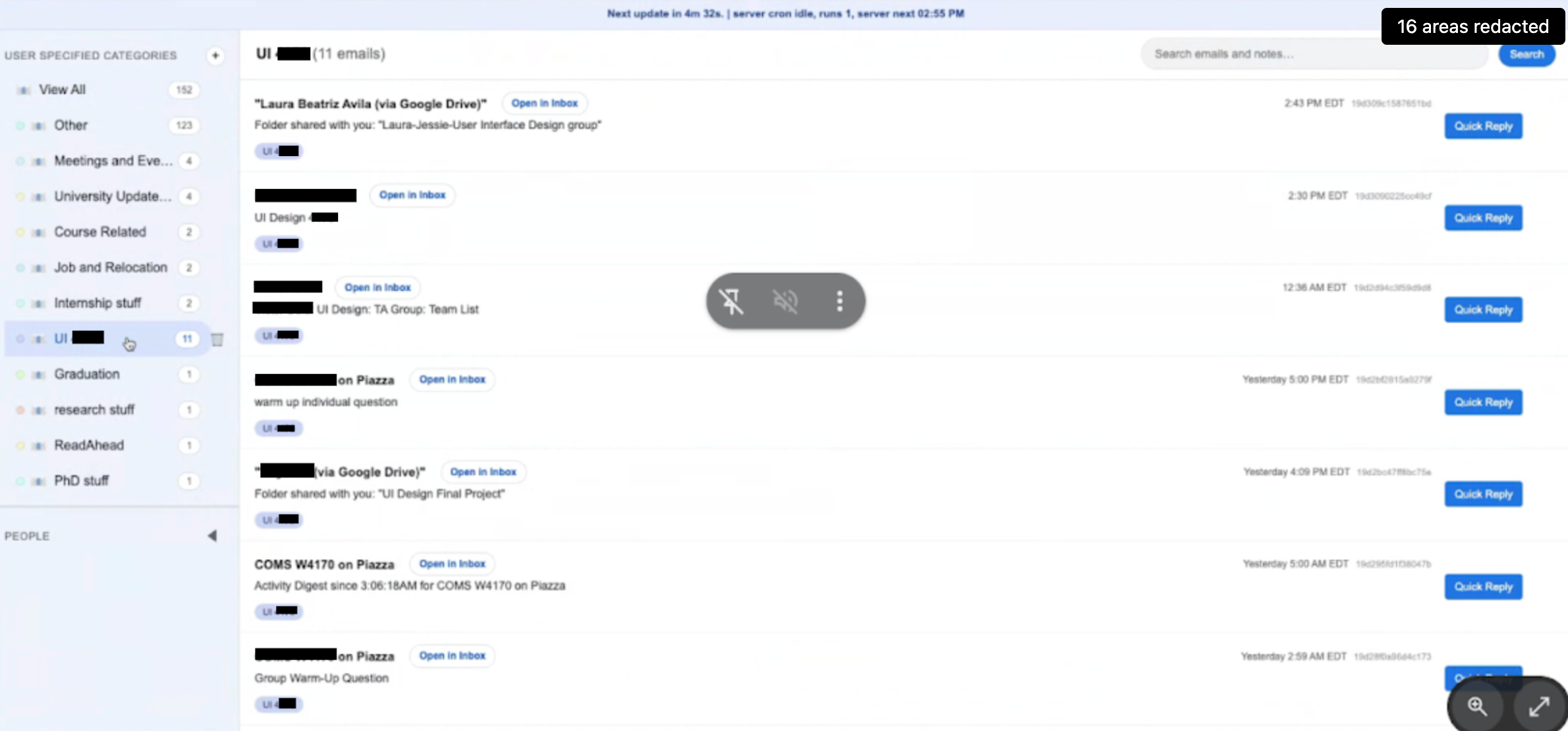}
    \caption{\textbf{Participant P4’s Day 1 feature introducing category-specific quick replies.} 
    P4 adds a ``Quick Reply'' button directly within emails belonging to a course-related category, enabling rapid responses to common student requests such as extension inquiries or questions already answered on the course website. This feature reduces the need to repeatedly compose similar replies by allowing responses to be generated in-place within the existing inbox interface.}
    \Description{An email inbox interface where each message in a course-related category includes a ``Quick Reply'' button for rapidly responding to common student inquiries.}
    \label{figures:day1p4}
\end{figure}

\begin{figure}[t]
    \centering
    \includegraphics[width=\linewidth]{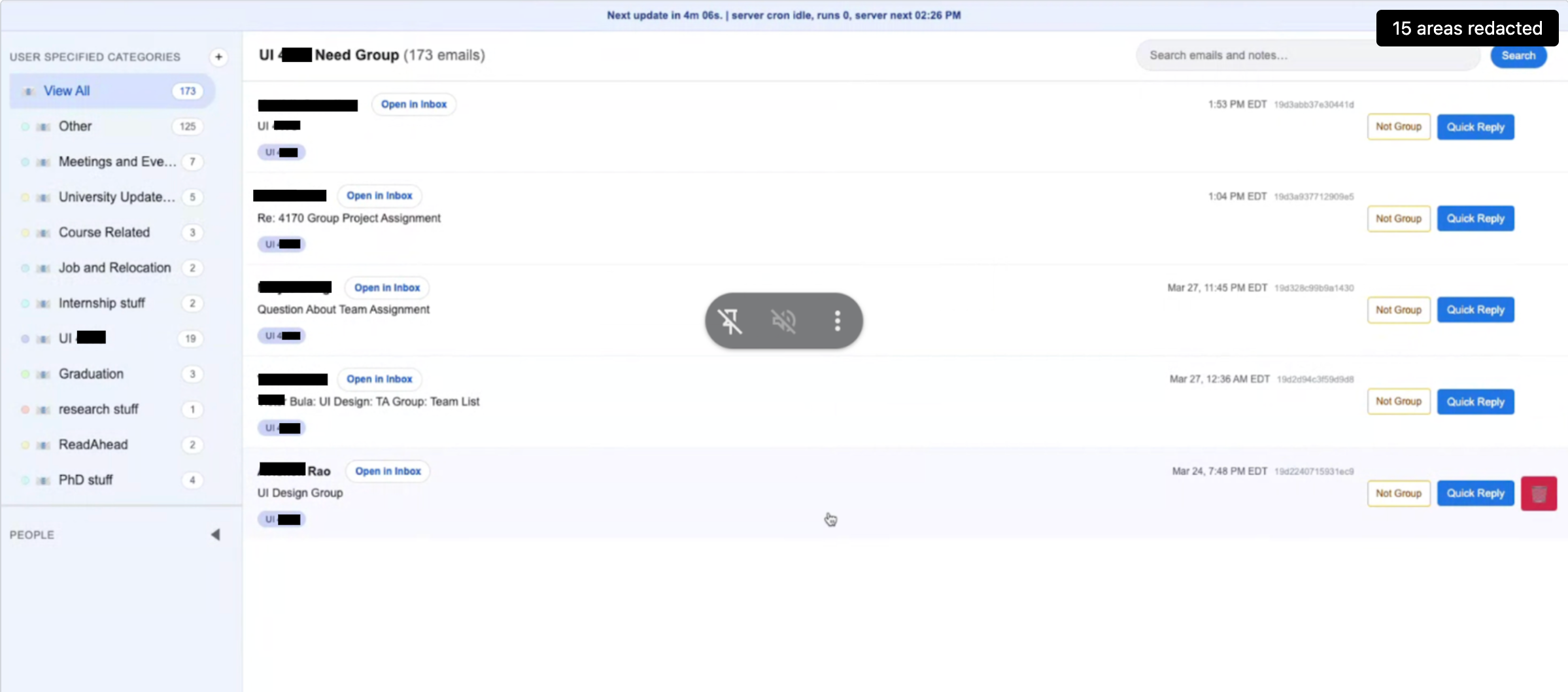}
    \caption{\textbf{Participant P4’s Day 2 iteration introducing filtering for group-related requests.} 
    Building on the quick reply functionality, P4 adds a filter that identifies emails from students requesting group formation. These emails are grouped into a dedicated view (``Need Group''), enabling the participant to isolate and manage coordination-related requests separately from other course communication.}
    \Description{An inbox view filtered to show emails categorized as group requests, labeled ``Need Group,'' with actions such as quick reply available for each message.}
    \label{figures:day2p4}
\end{figure}

\begin{figure}[t]
    \centering
    \includegraphics[width=\linewidth]{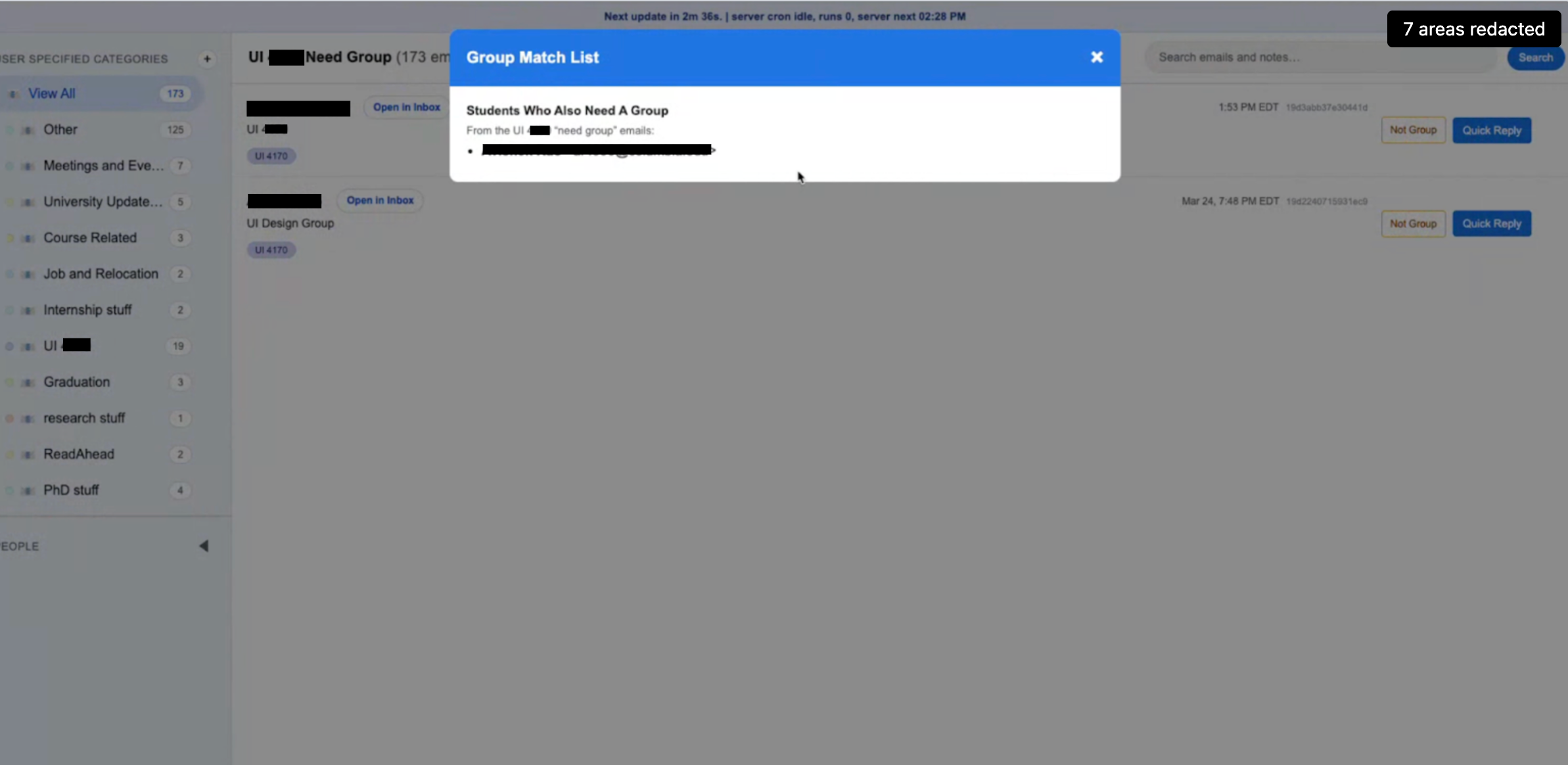}
    \caption{\textbf{Further iteration surfacing coordination information for group formation.} 
    From the filtered set of ``Need Group'' emails, the system extracts and aggregates students who are seeking group members, presenting them in a pop-up ``Group Match List.'' This allows P4 to quickly identify and connect students with similar needs, extending the feature from simple filtering to supporting coordination across multiple email threads.}
    \Description{A pop-up interface titled ``Group Match List'' displaying a list of students extracted from emails who are seeking group members.}
    \label{figures:day2p4_1png}
\end{figure}

\clearpage
\section{P5 | Design Probe}
\begin{figure}[H]
    \centering
    \includegraphics[width=\linewidth]{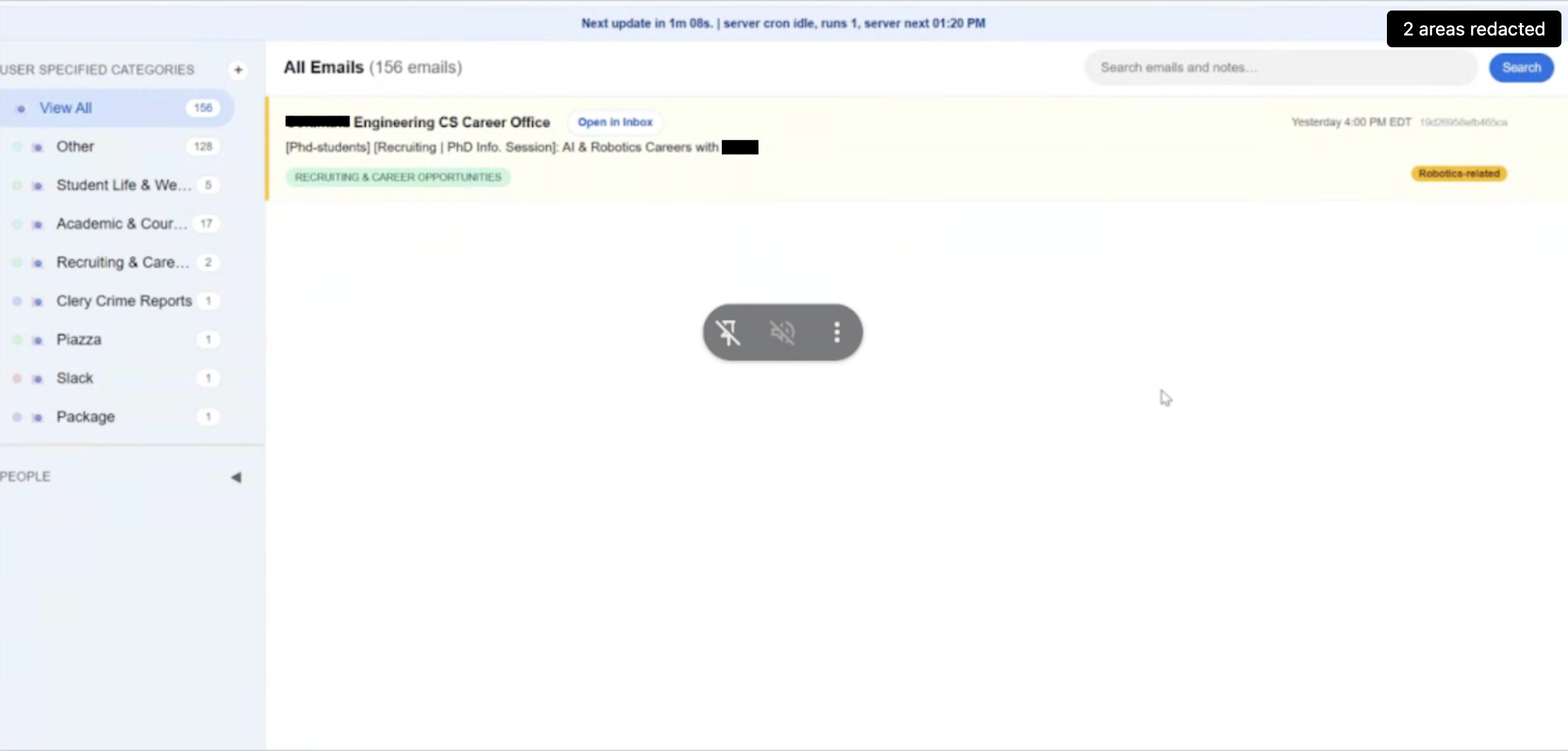}
    \caption{\textbf{Participant P5’s Day 1 feature for filtering and prioritizing robotics-related emails.} 
    P5 creates a filter that identifies emails related to robotics events and surfaces them within the inbox, allowing relevant opportunities to be distinguished from other incoming messages such as promotions or unrelated updates. Highlighted threads are visually emphasized, enabling quick identification without requiring manual scanning of the inbox. Unlike other participants, P5 did not iterate on this feature further, as the initial functionality was sufficient for their needs.}
    \Description{An email inbox interface where a robotics-related email thread is highlighted and prioritized among other messages, which include promotional and job application emails.}
    \label{figures:day1p5}
\end{figure}

\clearpage
\section{P6 | Design Probe}
\begin{figure}[H]
    \centering
    \includegraphics[width=\linewidth]{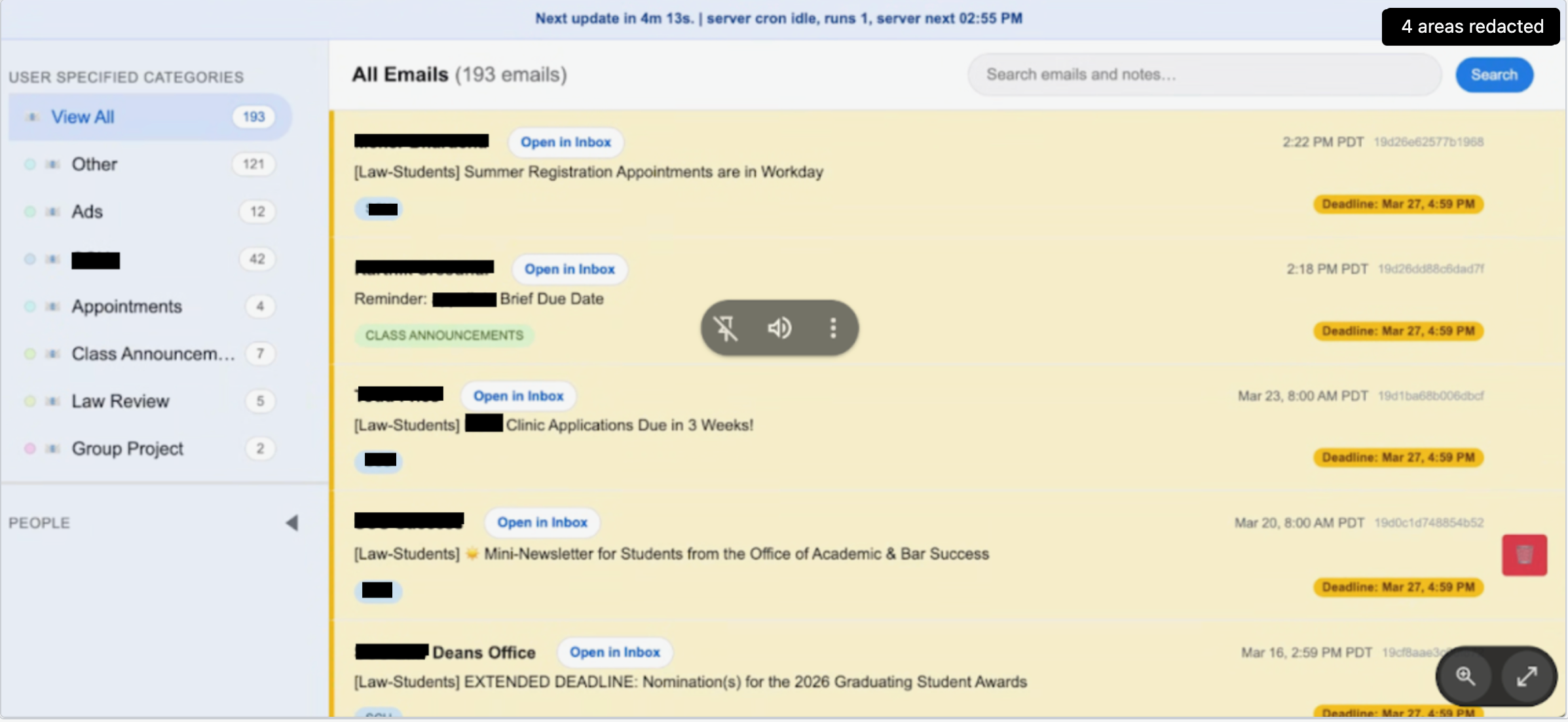}
    \caption{\textbf{Participant P6’s Day 1 feature surfacing deadline-related emails.} 
    P6 creates a feature that identifies emails containing deadlines and highlights them directly within the inbox. Extracted deadlines are displayed alongside each thread, allowing time-sensitive messages (e.g., applications, assignments, or announcements) to be quickly distinguished from other emails. This reduces the need to manually scan message content to identify urgent items.}
    \Description{An email inbox interface where certain emails are highlighted and annotated with extracted deadline information, indicating due dates associated with each message.}
    \label{figures:day1p6}
\end{figure}

\begin{figure}[t]
    \centering
    \includegraphics[width=\linewidth]{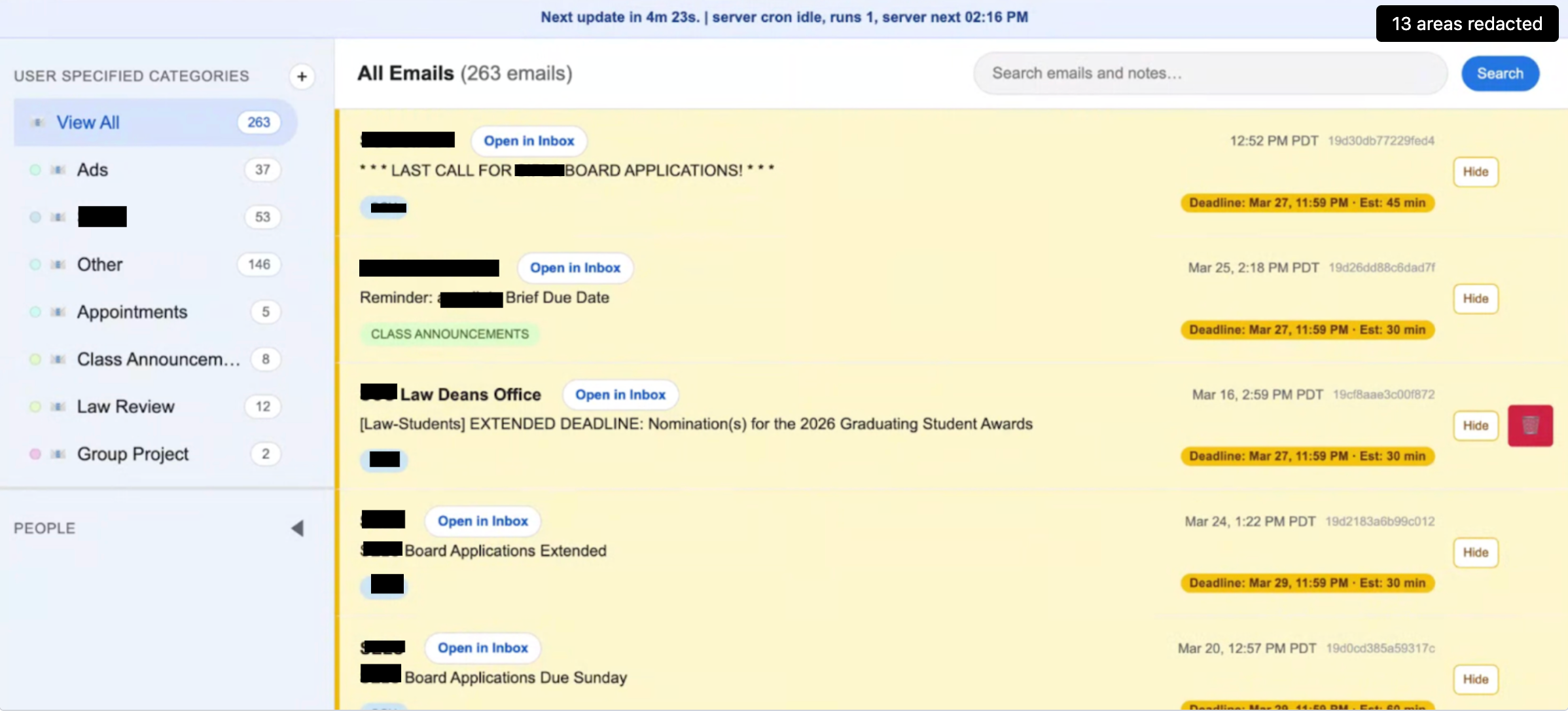}
    \caption{\textbf{Participant P6’s Day 2 iteration improving deadline accuracy and adding time estimates.} 
    Following initial use, P6 refines the feature to correct inaccuracies in extracted deadline times and augments each deadline with an estimated time remaining until it occurs. These additions improve both the reliability and usefulness of the feature, enabling more precise prioritization of time-sensitive emails based on urgency.}
    \Description{An updated inbox interface where emails display corrected deadline timestamps along with estimated time remaining until each deadline.}
    \label{figures:day2p6}
\end{figure}

\clearpage
\section{P7 | Design Probe}
\label{designprobe:p7}
\begin{figure}[H]
    \centering
    \includegraphics[width=\linewidth]{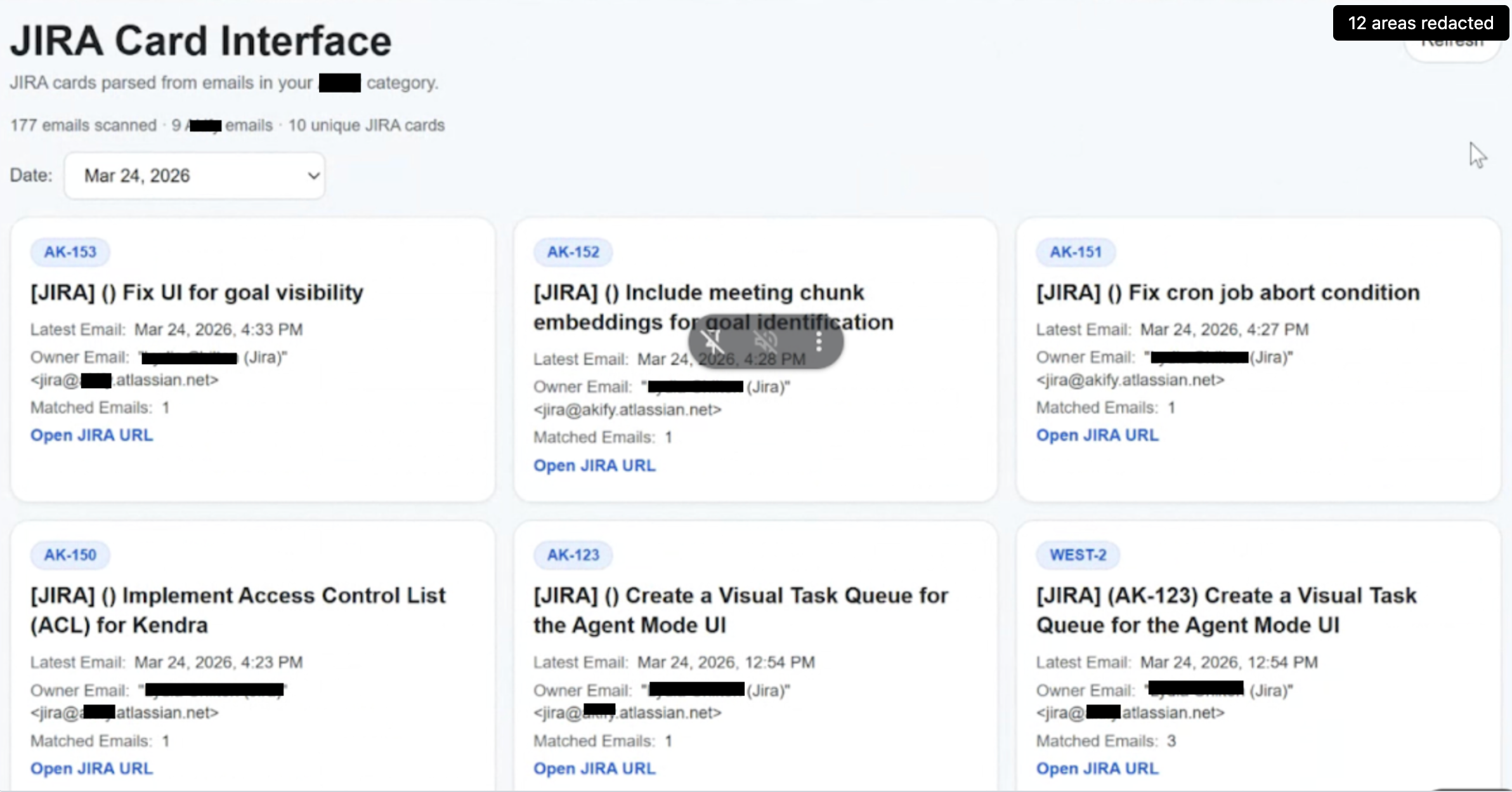}
    \caption{\textbf{Participant P7’s Day 1 feature constructing a JIRA-style interface from email data.} 
    P7 creates a separate interface that extracts and aggregates JIRA-related information from emails, presenting them as structured cards with fields such as task title, owner, latest update, and links to the corresponding JIRA issue. This derived view enables P7 to monitor project tasks without navigating across individual email threads, effectively transforming email notifications into a lightweight task management interface.}
    \Description{A dashboard interface displaying multiple JIRA-style cards generated from emails, each containing task titles, owners, timestamps, and links to corresponding issues.}
    \label{figures:day1p7}
\end{figure}
\begin{figure}[t]
    \centering
    \includegraphics[width=\linewidth]{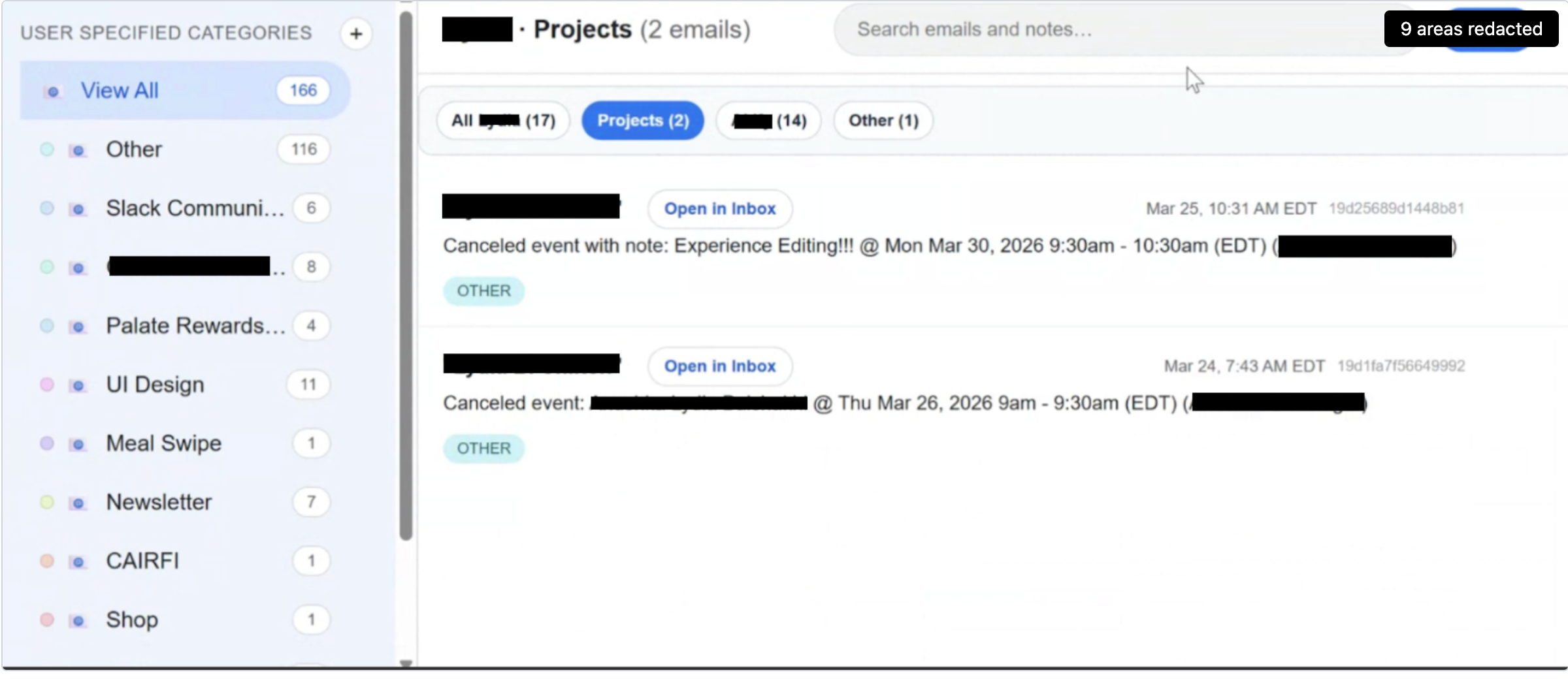}
    \caption{\textbf{Participant P7’s Day 2 iteration introducing categorical organization of advisor emails.} 
    Building on the initial interface, P7 restructures emails from her advisor into distinct categories corresponding to different types of work (e.g., research projects, startup-related tasks, and other communication). These categories are surfaced within the inbox sidebar, enabling more targeted navigation and organization of project-related emails. This iteration reflects a shift from a separate derived interface toward modifying the underlying inbox structure to better align with ongoing work contexts.}
    \Description{An email inbox interface where messages are organized into user-defined categories, including project-related groupings, shown in the sidebar and used to filter visible emails.}
    \label{figures:day2p7}
\end{figure}

\clearpage
\section{Day 3 Survey Results}

\subsection{Quantitative Measures}

Participants reported generally positive outcomes across measures of feature effectiveness, effort, and trust.

\vspace{0.5em}

\begin{table}[H]
\centering
\footnotesize
\renewcommand{\arraystretch}{1.1}
\begin{tabular}{p{0.55\linewidth} c c}
\toprule
\textbf{Measure} & \textbf{Mean} & \textbf{Median} \\
\midrule
\multicolumn{3}{l}{\textit{Feature outcome}} \\
\quad Addressed original motivation & 4.38 & 4 \\
\quad Behaved accurately & 3.75 & 3.5 \\
\quad Effort to create (1=low, 5=high) & 2.88 & 3 \\
\midrule
\multicolumn{3}{l}{\textit{Long-term adoption}} \\
\quad Likely to use in 1 month & 3.63 & 4 \\
\quad Would extend/modify further & 3.88 & 4 \\
\quad Saves time/effort & 3.63 & 4 \\
\midrule
\multicolumn{3}{l}{\textit{Risk \& trust}} \\
\quad Concerned about missing emails & 1.88 & 2 \\
\quad Confident I understand the feature & 4.38 & 4.5 \\
\quad Comfortable running unsupervised & 3.88 & 4 \\
\bottomrule
\end{tabular}
\caption{Day 3 structured survey results (1--5 Likert scale, $N{=}8$). Participants reported generally positive outcomes and moderate comfort with unsupervised feature behavior.}
\label{tab:survey}
\end{table}

\vspace{1em}

\subsection{Participant Reflections}

We also collected open-ended reflections on desired improvements, perceived risks, and factors influencing long-term adoption.

\vspace{0.5em}

\begin{table}[H]
\centering
\footnotesize
\renewcommand{\arraystretch}{1.1}
\begin{tabular}{p{0.12\linewidth} p{0.82\linewidth}}
\toprule
 & \textbf{What would you change?} \\
\midrule
P0 & Adding a visual globe to show where articles are referencing geographically. \\
P1 & Connect to additional data to improve ranking quality and visual design. \\
P2 & Better looking. \\
P3 & Displayed emails cleaned up further, fewer links, focused on article headers. \\
P4 & More robust, respond to wider set of TA questions by scraping other inbox information. \\
P5 & Better understanding of related fields to robotics without using keywords. \\
P6 & N/A. \\
P7 & Emails tagged more efficiently without manual effort. \\
\bottomrule
\end{tabular}
\caption{Participant responses describing desired improvements.}
\label{tab:changes}
\end{table}

\vspace{0.8em}

\begin{table}[H]
\centering
\footnotesize
\renewcommand{\arraystretch}{1.1}
\begin{tabular}{p{0.12\linewidth} p{0.82\linewidth}}
\toprule
 & \textbf{Worst-case if feature misbehaved?} \\
\midrule
P0 & Poor summarizations and fake tickers. \\
P1 & Missing apartment units could have financial cost or require manual verification. \\
P2 & Delete my emails. \\
P3 & Incorrect summaries---possibly jumbled outputs. \\
P4 & Give students incorrect information or come across as rude or cold. \\
P5 & A useless tab of emails. \\
P6 & Missing emails. \\
P7 & Miss an important meeting email. \\
\bottomrule
\end{tabular}
\caption{Participant responses describing worst-case failure scenarios.}
\label{tab:failures}
\end{table}

\vspace{0.8em}

\begin{table}[H]
\centering
\footnotesize
\renewcommand{\arraystretch}{1.1}
\begin{tabular}{p{0.12\linewidth} p{0.82\linewidth}}
\toprule
 & \textbf{What would increase regular use?} \\
\midrule
P0 & Integrating my logins into newsletter sites. \\
P1 & Likely short-term use tied to apartment search. \\
P2 & Nothing. \\
P3 & Direct links alongside summaries. \\
P4 & Responses should better match my tone before full automation. \\
P5 & Unlikely—I rarely use email features. \\
P6 & Running smoother. \\
P7 & More robust, bug-free. \\
\bottomrule
\end{tabular}
\caption{Participant responses describing factors influencing continued use.}
\label{tab:regularuse}
\end{table}

\vspace{1em}

\subsection{Cross-User Feature Interest}

Finally, we asked participants which features created by others they would consider using.

\vspace{0.5em}

\begin{table}[H]
\centering
\footnotesize
\renewcommand{\arraystretch}{1.1}
\begin{tabular}{p{0.70\linewidth} c}
\toprule
\textbf{Feature} & \textbf{Votes} \\
\midrule
Job application tracker & 5 \\
Deadline email highlighter & 5 \\
Bloomberg-style newsletter digest & 4 \\
Jira card viewer & 4 \\
Newsletter article tracker & 2 \\
Google Scholar reading list & 2 \\
Auto-reply student emails & 2 \\
Apartment listings ranker & 1 \\
Robotics talk filter & 1 \\
\bottomrule
\end{tabular}
\caption{Cross-user feature interest (multiple selections allowed, $N{=}8$). Broadly applicable features received the most interest.}
\label{tab:community}
\end{table}

\end{document}